\documentclass[preprint,11pt]{elsarticle}
\usepackage{graphicx}
\usepackage{verbatim}
\usepackage{epsfig}
\usepackage{amsmath}
\usepackage{amsfonts}
\usepackage{amssymb}
\usepackage{mathrsfs}
\usepackage{rotating}
\usepackage{mathtools}
\usepackage{multirow}
\usepackage{amsthm}
\usepackage{esvect}
\usepackage{multicol}
\usepackage{subcaption}
\usepackage{algorithm}
\usepackage{algorithmic}
\usepackage{enumerate}
\usepackage{bm}
\usepackage{hyperref}
\def\be{\begin{equation}}
\def\ee{\end{equation}}
\def\bq{\begin{eqnarray}}
\def\eq{\end{eqnarray}}

\newtheorem{theorem}{Theorem}[section]

\newtheorem{remark}[theorem]{Remark}

\makeatletter
\def\ps@pprintTitle{%
  \let\@oddhead\@empty
  \let\@evenhead\@empty
  \def\@oddfoot{\reset@font\hfil\thepage\hfil}
  \let\@evenfoot\@oddfoot
}
\makeatother

\usepackage{amssymb}

\begin{document}
\begin{frontmatter}
	
\title{Dynamic graph neural networks for enhanced volatility prediction in financial markets}
\author{Pulikandala Nithish Kumar\footnotemark[1]; Nneka Umeorah\footnotemark[2]; Alex Alochukwu\footnotemark[3]}
\address{\footnotemark[1]Cardiff University; School of Mathematics; Cardiff CF24 4AG; United Kingdom [pulikandalan@cardiff.ac.uk]\\
\footnotemark[2]Cardiff University; School of Mathematics; Cardiff CF24 4AG; United Kingdom [umeorahn@cardiff.ac.uk]\\
\footnotemark[3]Albany State University; Department of Mathematics, Computer Science and Physics; Albany, Ga 31705; USA [alex.alochukwu@asurams.edu]\\
\textbf{Correponding author}: Pulikandala Nithish Kumar [pulikandalan@cardiff.ac.uk] }

\begin{abstract}
Volatility forecasting is essential for risk management and decision-making in financial markets. Traditional models like Generalized Autoregressive Conditional Heteroskedasticity (GARCH) effectively capture volatility clustering but often fail to model complex, non-linear interdependencies between multiple indices. This paper proposes a novel approach using Graph Neural Networks (GNNs) to represent global financial markets as dynamic graphs. The Temporal Graph Attention Network (Temporal GAT) combines Graph Convolutional Networks (GCNs) and Graph Attention Networks (GATs) to capture the temporal and structural dynamics of volatility spillovers. By utilizing correlation-based and volatility spillover indices, the Temporal GAT constructs directed graphs that enhance the accuracy of volatility predictions. Empirical results from a 15-year study of eight major global indices show that the Temporal GAT outperforms traditional GARCH models and other machine learning methods, particularly in short- to mid-term forecasts. The sensitivity and scenario-based analysis over a range of parameters and hyperparameters further demonstrate the significance of the proposed technique. Hence, this work highlights the potential of GNNs in modeling complex market behaviors, providing valuable insights for financial analysts and investors.\\

\noindent\textbf{Keywords}: GARCH model $\cdot$ Graph Neural network $\cdot$ Temporal GAT $\cdot$ Volatility cluster $\cdot$ Optimization $\cdot$ Volatility spillover.\\
\textbf{JEL}:  C15 \sep C45 \sep C53 \sep C58 \sep  G17.
\end{abstract}

\end{frontmatter}


\section{Introduction}\label{sec1}

\noindent Volatility in stock markets refers to the degree of variation in asset prices over time and plays a critical role in assessing risk and uncertainty in financial markets. One of the most perplexing aspects of volatility is volatility clustering, a phenomenon where periods of large price changes are followed by further large changes, and periods of small changes are followed by small changes. This clustering behaviour, well-documented phenomenon in financial markets\citep{engle1993statistical}, significantly influences the decision-making processes of investors and risk managers as they adjust their portfolios based on the perceived level of market risk and uncertainty. The clustering patterns are often obvious during market turbulence or uncertainty and could be attributed to market participants reacting to new information, whether economic or financial shocks, news events, or changing market sentiment\citep{cont2001empirical}. During such events, volatility rises as uncertainty increases, leading to further extreme price movements before stabilizing. \\

\noindent This concept was first observed by \citet{mandelbrot1972certain} in 1963 and further developed by \citet{engle1982autoregressive} in 1982 with the introduction of the Autoregressive Conditional Heteroskedasticity (ARCH) model. Engle's work demonstrated that volatility is not constant over time but can be modeled based on past behaviors, marking a major shift in understanding financial time series. Building on this, Bollerslev extended the ARCH model by proposing the Generalized Autoregressive Conditional Heteroskedasticity (GARCH) model in 1986\citep{bollerslev1986generalized}. The GARCH model improved upon its predecessor by incorporating past volatilities in addition to past shocks, allowing for more accurate forecasting of future volatility. Traditionally, models like the GARCH model have been widely used to capture volatility clustering by analyzing time-series data. While effective to an extent, these models often fail to account for the more complex, non-linear relationships between multiple assets or even networks of interdependent assets in global financial markets\citep{nelson1991conditional}. This limitation becomes particularly evident when markets are highly interconnected, and the volatility of one market affects others through intricate spillover effects\citep{diebold2012better}.\\

\noindent Several variations of the GARCH model have been developed to account for specific market behaviors. For instance, the Exponential GARCH (EGARCH) model introduced by Nelson in 1991 addresses asymmetries in financial data, particularly the tendency for negative market shocks to increase volatility more than positive ones\citep{nelson1991conditional}. Similarly, the Threshold GARCH (TGARCH) model proposed by Zakoian in 1994 captures the idea that markets respond differently to various types of shocks\citep{zakoian1994threshold}. Despite their widespread adoption, GARCH models have notable limitations. They often rely heavily on historical data, which can be problematic in rapidly evolving markets. Moreover, they assume that volatility follows a single process, which overlooks the complex, interconnected relationships between global financial markets\citep{diebold2012better}. These challenges have led to exploring alternative approaches, including machine learning techniques, to enhance volatility forecasting. Recent advancements have seen the integration of realized volatility measures to capture the dynamic nature of financial markets better. Andersen et al. (2013) demonstrated the effectiveness of realized volatility in providing more accurate volatility estimates by utilizing high-frequency intraday data\citep{andersen2013financial}. Additionally, the concept of volatility spillovers, as measured by the Volatility Spillover Index, has been instrumental in understanding how volatility in one market can influence others.\\

\noindent To model these interdependencies, the Graph Neural Network (GNN), a class of deep learning techniques, has been designed to leverage and operate on specified graph-structured data. GNNs provide a sophisticated way to model time-series patterns and the relationships between different market indices by treating them as nodes within a graph structure\citep{scarselli2008graph}. Each edge in this graph represents a relationship or correlation between indices, capturing the dynamic interdependencies between global markets. Traditional machine learning applications handle graph-structured data by using a preprocessing step that transforms the graph's structured information into a more straightforward form, such as real-valued vectors. However, crucial information, such as the topological relationships between nodes, can be lost during the preprocessing stage. As a result, the final outcome may be unpredictably influenced by the specifics of the preprocessing algorithm\citep{haykin1998neural}. Other approaches that have attempted to preserve the graph-structured nature of the data before the processing phase have been investigated by other researchers in \citep{frasconi1998general},\citep{scarselli2008graph}, \citep{brin1998anatomy}.\\

\noindent In recent years, GNNs have gained attention, especially in financial modelling and research, due to their capacity to capture both structural and temporal dependencies evident in a financial system. This ability to account for both the structural topology and temporal dynamics makes GNNs a powerful tool for tasks like risk assessment, fraud detection, stock price prediction, and portfolio optimization in modern finance. Several comprehensive reviews on GNNs have been published. \citet{bronstein2017geometric} offers an in-depth review of geometric deep learning, discussing key challenges, solutions, applications, and future directions. Similarly, \citet{zhang2019graph} provides a detailed overview of graph convolutional networks, whereas \citet{zhou2020graph} investigated other computation modules in GNNs, such as skip connections and pooling operators. Other recent papers on GNN models are presented by other researchers in \citep{chami2022machine}, \citep{wu2020comprehensive}, \citep{sawhney2020deep} and  \citep{zhang2020deep}.\\

\noindent Graph-based models, such as the Temporal Graph Attention Networks (GATs), have shown great promise in learning from structured data, particularly graphs. The model enhances the traditional methods of analyzing financial data by incorporating both graph convolutional networks (GCNs) and GATs\citep{kipf2016semi}. This allows the model to focus on the most relevant relationships between indices, improving the accuracy of predictions related to volatility clustering. Unlike traditional models that only consider individual asset behaviors, the Temporal GAT model enables a deeper understanding of how assets influence each other, providing a more comprehensive analysis of market volatility. GATs utilize attention mechanisms to assign varying importance to nodes within a graph, making them practical for modeling relationships between financial entities. However, traditional GATs struggle to capture temporal dependencies in data, which is crucial in the financial domain where time plays a key role.\\

\noindent To address this limitation, Temporal GATs have been developed to enhance GATs by incorporating temporal data. Using a temporal attention mechanism, the temporal GAT dynamically adjusts the impact of past events, allowing the model to capture not only the structure of the financial graph but also the time-dependent patterns that shape market behavior\citep{xu2020inductive},\citep{zhou2020graph},\citep{xiang2022temporal}. This phenomenon makes the temporal GATs particularly effective for financial applications, where it is crucial to consider both entity relationships and their changes over time. This research aims to push the boundaries of financial econometrics and machine learning by applying GNNs to understand volatility clustering better and predict it. By capturing the interconnected nature of global financial markets, this paper offers new tools for financial analysts, risk managers, and investors to improve risk assessment and make more informed decisions\citep{sirignano2021universal}. \\

\noindent The key contributions of this paper are as follows:

\begin{itemize}
	\item \textbf{Development of a Temporal GAT for Volatility Forecasting:} This paper presents the design and application of a novel Temporal GAT model tailored for volatility prediction in global stock markets. The model effectively captures both temporal dependencies and interdependencies between global market indices, outperforming traditional methods like GARCH and MLP.
	
	\item 	\textbf{Volatility Spillover Index as a Superior Graph Construction Method:} The research demonstrates the effectiveness of using a volatility spillover index to model the relationships between market indices. This approach significantly enhances the model's ability to capture the transmission of market shocks compared to conventional correlation-based methods.
	
	\item 	\textbf{Comprehensive Sensitivity Analysis and Robustness Testing:} The study provides an in-depth sensitivity analysis and tests the model's robustness across different market conditions, including high and low volatility periods. These analyses help identify the key factors influencing model performance and offer guidance for applying the Temporal GAT model in various financial environments.
\end{itemize}

\noindent The practical significance of this research lies in its ability to capture and model complex interdependencies between multiple indices, enhance volatility predictions, and optimize risk management techniques. Traditional models like GARCH focus on each asset's volatility but fail to account for how volatility in one asset can spill over to other assets and vice versa. GNNs address this limitation by representing assets as nodes in a graph, with edges reflecting relationships such as correlations or mutual volatility impacts. This allows GNNs to capture interconnected market behaviour better, providing more accurate volatility clustering models. Some of the practical benefits are seen in modelling complex asset relationships, cross-asset and multi-asset volatility spillovers, real-time systematic risk monitoring in terms of financial contagion, algorithmic trading and portfolio optimization in terms of optimizing risk-adjusted returns and handling correlation breakdowns in the face of market turmoil. \\

\noindent The rest of the paper is organized as follows: Section \ref{sec2} introduces the literature studies in terms of volatility clustering, machine learning approaches to volatility forecasting, graph neural networks (GNN) in financial modelling and the concepts of volatility spillovers and the GNNs. Section \ref{sec3} describes the mathematical concepts and the preliminaries of the research, and section \ref{sec4} introduces the methodology. Section \ref{sec5} discusses the empirical results, together with the data visualization, sensitivity and robustness analysis, and the last section concludes the work.

\section{Related Works}\label{sec2}
\noindent Regarding the Machine learning (ML) techniques for volatility forecasting, ML has gained prominence in financial volatility forecasting due to its ability to model complex, non-linear relationships. Neural networks, such as Long Short-Term Memory (LSTM) networks, have been applied to volatility forecasting with notable success\citep{kim2018forecasting, liu2019novel}. These deep learning models improve upon traditional approaches like GARCH by capturing long-term dependencies in financial data, which is critical for accurate volatility prediction. The combination of machine learning and GARCH as a hybrid model has been effective in volatility forecasting and is applicable in various markets such as energy, main metals, and especially stock markets\citep{amirshahi2023hybrid}. Artificial Neural Networks (ANNs), Support Vector Machines (SVMs), and Random Forests have also been used in financial forecasting with varying success. ANNs effectively capture non-linear patterns in the data, but they require large datasets and can struggle with overfitting\citep{kumar2014forecasting, kurani2023comprehensive}. SVMs have been applied to classify market conditions, and random forests have provided robust results, particularly in large datasets. However, both models typically treat assets independently, failing to capture the interconnectedness of global markets.\\

\noindent On the other hand, the GNNs have emerged as a powerful tool for modeling complex, interconnected systems like financial markets. \citet{xu2018powerful} presented a theoretical framework for analyzing the representational power of GNNs, as well as their variants. GNNs can capture spatial and temporal dependencies, making them ideal for forecasting volatility across global markets. \citet{son2023forecasting} recently demonstrated how GNNs, combined with a volatility spillover index, can enhance the predictive accuracy for stock market volatility across different regions\citep{son2023forecasting}. The foundational work on GNNs can be traced to \citet{scarselli2008graph}, who introduced a framework for applying neural networks to graph-structured data. Their model allowed nodes to iteratively update their representations based on information from their neighbours, laying the groundwork for more advanced GNN architectures\citep{scarselli2008graph}.\\

\noindent Graph Convolutional Networks (GCNs), introduced by Kipf and Welling in 2016, were a significant advancement in applying GNNs to financial modelling\citep{kipf2016semi}. GCNs aggregate information from neighbouring nodes to update node features based on the graph's topology, making them highly effective for semi-supervised learning on graphs. GCNs can incorporate temporal dynamics, enabling them to model changes over time, which is crucial in the fast-paced financial markets\citep{li2018deeper}. \citet{yin2021forecasting} further combined GCN and gated recurrent unit (GRU) to forecast stock prices using a stock correlation graph. They utilized the GCN to extract features from the price of each stock price, and this sequence of features is then fed into a GRU model to capture temporal dependence. Other recent works on the implementation of the GCNs to financial modelling are found in \citep{chen2021novel},\citep{chen2018incorporating}, \citep{alarab2023graph} and \citep{ye2021multi}. On the other hand, GATs were developed by \citet{velivckovic2017graph} in 2018, and they enhance GCNs by incorporating attention mechanisms that dynamically assign different weights to neighbouring nodes, allowing the model to focus on the most relevant connections. This innovation has made GATs particularly useful in applications in multivariate time series prediction \citep{wang2022network}, financial fraud detection \citep{wang2019semi} and in other financial modeling framework, where some market indices have a more significant influence on others.\\

\noindent Finally, connecting volatility spillovers with GNNs, Diebold and Yilmaz (2009, 2012) developed a framework to measure these spillovers using Vector Autoregressive (VAR) models\citep{diebold2009measuring}, \citep{diebold2012better}. Volatility spillovers refer to the phenomenon where volatility in one market influences volatility in another. Traditional VAR-based spillover measures may fail to capture non-linear relationships, which is where GNNs can provide a more nuanced understanding. Recent research has sought to overcome these limitations by using GNNs to model the global financial system as a dynamic graph\citep{kipf2016semi}. GNNs can provide a more comprehensive understanding of volatility spillovers by representing markets as nodes and volatility relationships as edges. \citet{son2023forecasting} demonstrated that incorporating volatility spillover indices into a spatial-temporal GNN framework significantly improves the accuracy of volatility predictions across global markets.

\section{Preliminaries} \label{sec3}

This section introduces the fundamental concepts and methodologies essential for understanding the subsequent analysis of volatility clustering using GNNs. The topics covered include realized volatility, the GARCH model, correlation, the volatility spillover index, and the architectures of GCNs and GATs. These concepts form the backbone of the proposed Temporal GAT model.

\subsection{Volatility and Correlation Analytics}

\subsubsection{Realized Volatility}
Realized volatility is a non-parametric measure of the actual volatility observed in financial markets over a specific period. It is calculated by summing the squared returns over high-frequency intraday intervals, providing a more accurate reflection of market volatility than traditional measures. Mathematically, the realized volatility $(RV)$ for asset $i$ on day $t$ is given by:
\begin{align}
	RV_{i,t} &=\displaystyle \sum^{M}_{k=1}r^2_{t,k}
\end{align}

\noindent where $M$ represents the number of intraday intervals, and $rt$ denotes the return in the $k$-th interval of day $t$. Using high-frequency data enhances the estimation accuracy of volatility, capturing the intra-day price movements that daily closing prices might ignore.

\subsubsection{Generalized Autoregressive Conditional Heteroskedasticity (GARCH)}

The GARCH model, introduced by \citet{bollerslev1986generalized}, is a widely used statistical model for estimating volatility in financial time series data. It extends the ARCH model by incorporating both past squared returns and past variances\citep{zakoian1994threshold}, allowing for a more flexible and accurate modelling of volatility clustering \citep{diebold2012better}.

\noindent Given $\epsilon_t$ as a real-valued discrete-time stochastic process, and $f_t$ as the information set ($\sigma-$field) of all information through time $t$, then the standard GARCH $(p,q)$ model is defined as:

\begin{align}
	r_{t} &=\displaystyle \mu + \epsilon_{t}, \quad \epsilon_{t}|f_{t-1}\sim N(0,h_{t})\\
	h_{t} &=\displaystyle \omega_0 + \sum^{p}_{i=1} \alpha_{i} \epsilon^2_{t - i} + \sum^{q}_{j=1}\beta_{j}h_{t-j}
\end{align}

\noindent where\footnote{Note: For $q=0$, the process reduces to the ARCH($p$) process and for $p=q=0$, then $\epsilon_t$ becomes the white noise.} 
\[
  \begin{cases}
               q \geq 0, \; p>0\\
               \omega_0 >0; \; \alpha_i \geq 0, \; \text{for} \; i=1,2,\cdots p\\
               \beta_j \geq 0; \; \text{for} \; j=1,2,\cdots q\\
            \end{cases}
\]
\noindent and 
\begin{itemize}
	\item $r_t$ is the return at time $t$,
	\item $\mu$ and $\epsilon_t$ are the mean return and the error term, respectively.
	\item $h_t$ is the conditional variance (volatility) at time $t$
	\item $\omega , \alpha ,$ and $\beta$ are parameters to be estimated.
\end{itemize}

The GARCH model captures volatility clustering by allowing the current variance to depend on both past squared errors $(\epsilon^2_{t - 1})$ and past variance $ (h_{t - 1}) $.

\subsubsection{Correlation}

Correlation measures the linear relationship between two variables, indicating how changes in one variable are associated with changes in another. In financial markets, correlation between asset returns is crucial for portfolio diversification and risk management. The Pearson correlation coefficient between two assets $i$ and $j$ is calculated as:

\begin{align}
	\rho_{ij} &=\displaystyle \frac{Cov(r_{i}, r_{j})}{\sigma_{i}\sigma_{j}}
\end{align}

\noindent where:
\begin{itemize}
	\item $Cov(r_{i}, r_{j})$ is the covariance between returns $r_{i}$ and $r_{j}$
	\item $\sigma_{i}$ and $\sigma_{j}$ are the standard deviations of $r_{i}$ and $r_{j}$ respectively.
	
\end{itemize}

A correlation coefficient of $+$ 1  indicates a perfect positive linear relationship, $-$ 1  indicates a perfect negative linear relationship, and 0 implies no linear relationship.

\subsubsection{Volatility Spillover Index}

The Volatility Spillover Index measures the extent to which volatility shocks can transfer from one market to another, reflecting the interconnectedness of the global financial market. Diebold and Yilmaz (2009, 2012) developed a framework using variance decompositions from the VAR models to quantify these spillovers \citep{diebold2009measuring}, \citep{diebold2012better}.

The spillover index is calculated using the forecast error variance decompositions from the VAR model\citep{diebold2009measuring}:

\begin{align}
	\theta^{g}_{ij}(H) &=\displaystyle \frac{\sigma^{-1}_{jj} \sum_{h = 0}^{H-1}\left( e'_{i} A_{h}\sum e_{j} \right)^2 }{\sum_{h=0}^{H-1}\left( e'_{i} A_{h} \sum A'_he_{i} \right) }
\end{align}

\noindent where \begin{itemize}
	\item $H$ is the forecast horizon,
	\item $\sigma_{jj}$ is the standard deviation of the error term for variable $j,$
	\item $e_{i}$ is the selection vector with one at the $i-th$ position and zeros elsewhere,
	\item 	$A_h$ is the coefficient matrix at lag $h,$
	\item $\sum $ is the covariance matrix of the error terms.
\end{itemize}

The total spillover index is then given by\footnote{Note: To calculate the spillover index using the information available in the variance decomposition matrix, each matrix entry is normalized by
\[\tilde{\theta}^{g}_{ij}(H) = \frac{\theta^{g}_{ij}(H)}{\sum_{j=1}^N \theta^{g}_{ij}(H)}\]}

\begin{align}
	S^{g}(H) &=\displaystyle \frac{\sum_{i,j=1; i \neq j}^N \tilde{\theta}^{g}_{ij}(H)}{\sum_{i,j=1}\tilde{\theta}^{g}_{ij}(H)} \times 100 = \frac{\sum_{i,j=1; i \neq j}^N \tilde{\theta}^{g}_{ij}(H)}{N} \times 100
\end{align}
This index provides insights into how volatility in one market influences others, which is essential for understanding systemic risk and market dynamics.

\subsection{Graph-based Deep Learning Models}
\subsubsection{Graph Neural Networks (GNNs)}
GNNs are a class of neural networks designed to operate on graph-structured data, capturing dependencies among nodes via message passing between the nodes of graphs. GNNs are particularly useful for modelling relational data and have been successfully applied in various domains, including social networks, recommendation systems, and financial markets.

\subsubsection{Graph Convolutional Networks (GCNs)}
GCNs extend the concept of convolutional neural networks to graph data. They aggregate feature information from a node's neighbours to compute its new representation \citep{kipf2016semi}. The layer-wise propagation rule for a multilayer GCN is given by:
\begin{align}
	H^{l+1} &=\displaystyle \sigma\left( \tilde{D}^{-\frac{1}{2}}\tilde{A}\tilde{D}^{\frac{-1}{2}}H^{(l)}W^{(l)} \right) 
\end{align}

where: \begin{itemize}
	\item 	$H^{(l)}$ is the feature matrix at layer $l$; $H^{(l)} \in \mathbb{R}^{N \times D}$,
	\item   $\tilde{A} = A + I_N$ is the adjacency matrix of the undirected graph $\mathcal{G}$ with added self-loops, and $I_N$ is the identity matrix,
	\item 	$\tilde{D}_{ii} = \sum_j \tilde{A}_{ij}$,
	\item $W^{(l)}$ is the layer-specific trainable weight matrix,
	\item 	$\sigma$ is an activation function (e.g., ReLU) \citep{kipf2016semi}
\end{itemize}
GCNs effectively capture local neighbourhood structures in graphs, making them suitable for semi-supervised learning tasks on graph-structured data.

\subsubsection{Graph Attention Networks (GATs)}

GATs introduce an attention mechanism to GNNs, allowing the model to assign different importance weights to different nodes in a neighbourhood. The core idea is to compute attention coefficients $\sigma_{ij}$  that indicate the importance of node $ j's$ features to node $i$ \citep{bahdanau2014neural, velivckovic2017graph}.
\begin{align}
		\vec{h}_{i}'&=\displaystyle \sigma \left( \sum_{j \in \mathcal{N}_i} \alpha_{ij}{\bf{W}}\vec{h}_{j}  \right) 
\end{align}

\begin{align}
	\alpha_{ij} &=\displaystyle \frac{\exp\left( \text{LeakyReLu} \left( \vec{{\bf{a}}}^T\left[ {\bf W}\vec{h_{i}}|| {\bf W} \vec{h_{j}} \right]  \right)  \right) }{\sum_{k \in \mathcal{N}_i} \exp\left( \text{LeakyReLu} \left( \vec{{\bf{a}}}^T\left[ {\bf W}\vec{h_{i}}|| {\bf W} \vec{h_{k}} \right]  \right)  \right) }
\end{align}
%
where: 
\begin{itemize}
	\item ${\bf h} = {\vec{h_1}, \vec{h_2}, \cdots, \vec{h_N}}$; $\vec{h_i} \in \mathbb{R}^F$ are the input, with $N$= number of nodes and $F$, the number of features in each node.
	\item ${\bf W} \in \mathbb{R}^{F' \times F}$ is the weight matrix,
	\item $a: \mathbb{R}^{F'} \times \mathbb{R}^{F'} \rightarrow \mathbb{R}$ is the attention mechanism's weight vector.
	\item $||$ denotes concatenation and ${.}^T$ is the transposition.
	\item $\mathcal{N}_i$ is the set of neighborhood of node $i$ in the graph.
\end{itemize}

GATs enhance the model's capacity to focus on the most relevant parts of the graph, improving performance on tasks where relationships between nodes vary in importance.

\subsubsection{Temporal Graph Attention Network (Temporal GAT)}
The Temporal GAT combines the strengths of GCNs and GATs to model both the structural and temporal dynamics in graph-structured data \citep{de2018advances}. It is particularly effective for time-dependent graphs where node relationships evolve over time. In the context of financial markets, the Temporal GAT can capture how the influence between different market indices changes over time, which is crucial for modelling volatility clustering and spillovers.

The Temporal GAT processes the data through multiple layers:
\begin{enumerate}
	\item GCN Layers: Capture the initial structural information by aggregating features from neighbouring nodes.
	
	\item GAT Layers: Apply attention mechanisms to focus on the most influential nodes, allowing the model to weigh the importance of different relationships.
	
	\item Temporal Layers: Incorporate time-series data to model the evolution of node features and relationships over time.
\end{enumerate}
Combining these layers enables the Temporal GAT to effectively model complex temporal and spatial dependencies in financial data, improving the accuracy of volatility predictions.

\section{Methodology}\label{sec4}

\noindent This section outlines the comprehensive methodology adopted to explore and analyze volatility clustering in global stock markets using a Temporal GAT approach. The research aims to advance the understanding and forecasting of volatility by leveraging financial markets' dynamic and interconnected nature. A systematic, step-by-step approach is presented, providing clarity on the strategies, techniques, and tools employed throughout the study, ensuring transparency and replicability of the findings.

\subsection{Problem Formulation}
Volatility clustering is a phenomenon in financial markets where large price changes are likely to be followed by large changes, and small changes tend to be followed by small changes. Traditional models, such as the GARCH model, have effectively analyzed time-series data but often fail to capture complex, non-linear interactions across different market indices\citep{zakoian1994threshold}. This limitation becomes particularly significant in today's highly interconnected global financial markets, where volatility in one market can influence others through intricate spillover effects.\\

\noindent To address these shortcomings, this research introduces an innovative approach by modelling the stock market as a dynamic graph, where each market index is represented as a node interconnected with others, reflecting the complex real-world relationships within financial markets. By adopting this approach, the research seeks to:

\begin{itemize}
	\item Capture Volatility Clustering: Utilize GNNs to model the complex patterns of volatility clustering that traditional models might miss during periods of market turbulence.
	
	\item 	Learn from Temporal and Structural Data: Integrate both the temporal sequence of stock prices and the structural relationships between market indices into a cohesive learning model.
	
	\item 	Forecast Future Volatility: Develop a predictive model capable of forecasting future market volatilities by leveraging historical price behaviours and the interconnections between market indices.
	
	\item Adapt to Market Dynamics: Investigate how dynamic graph structures can reflect real-time market changes, enabling the model to adapt to evolving conditions effectively.
\end{itemize}

\subsection{Graph Construction}
The core of our methodology lies in representing the stock market as a directed graph, where the indices are nodes and the relationships between them form directed edges. In a directed graph, each edge has a direction, indicating the flow of influence or information from one node (market index) to another. This structure is particularly suitable for capturing the asymmetric relationships often observed in financial markets, where one market can significantly impact another without necessarily experiencing a reciprocal effect\citep{diebold2009measuring}. We employed two primary methods for constructing these directed graphs: the Correlation Method and the Volatility Spillover Index Method.\\

\begin{figure}[H]
	\centering
	\begin{subfigure}{0.3\textwidth}
		\centering
		\includegraphics[width=\linewidth]{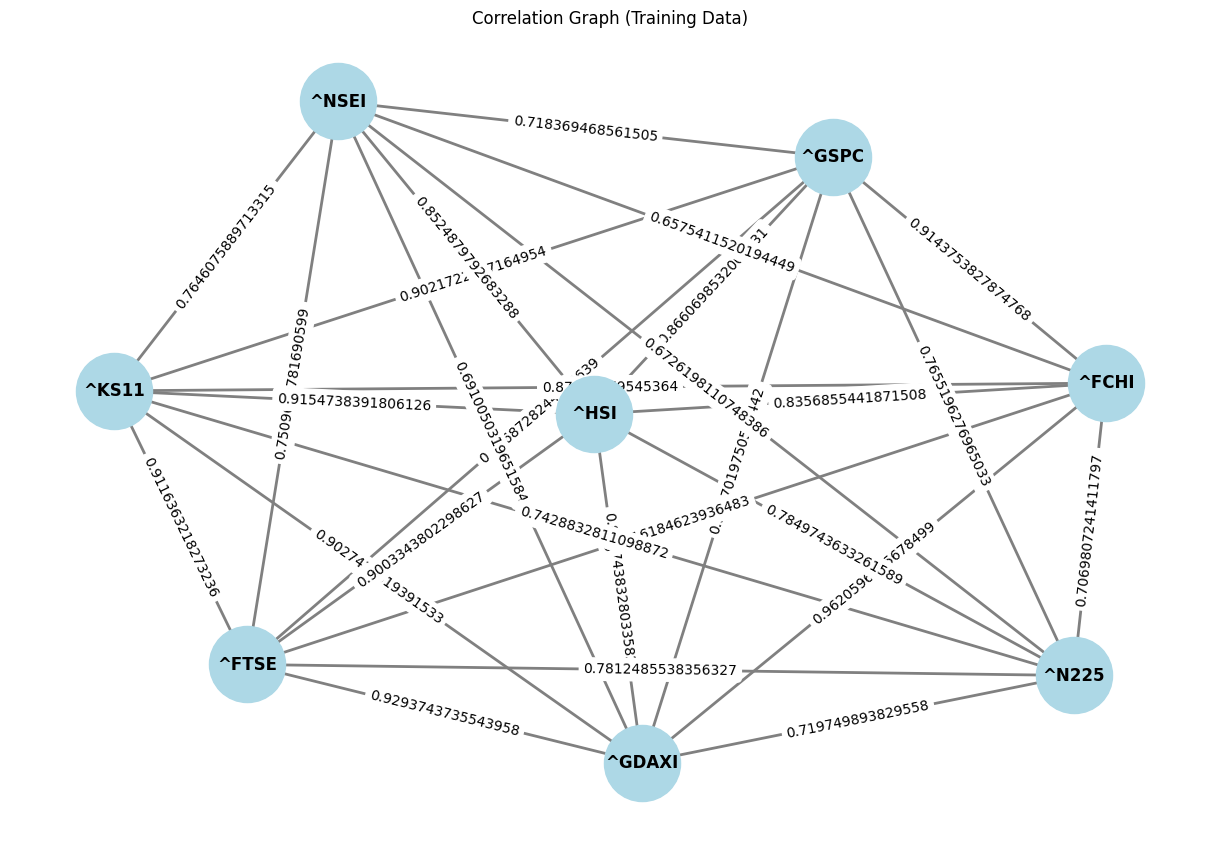}
	\end{subfigure}
	\hfill
	\begin{subfigure}{0.3\textwidth}
		\centering
		\includegraphics[width=\linewidth]{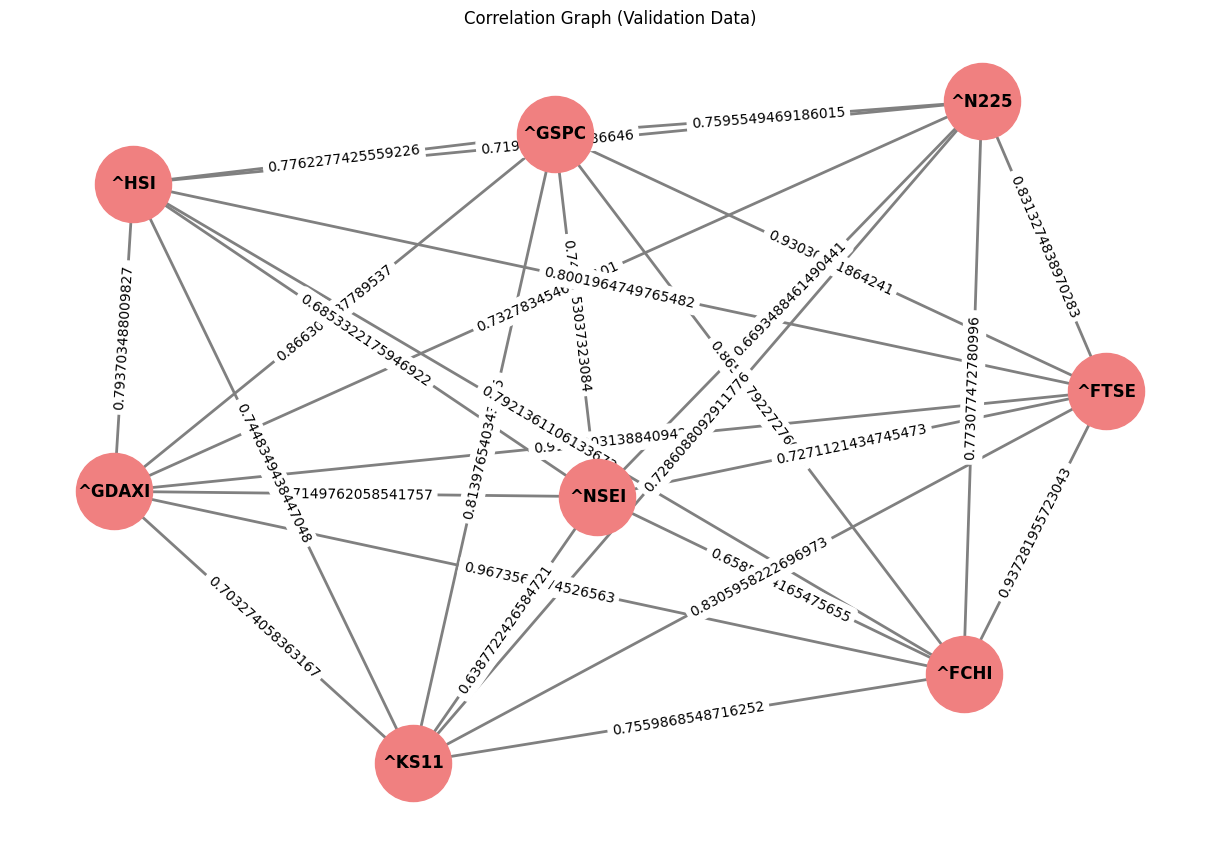}
		
	\end{subfigure}
	\hfill
	\begin{subfigure}{0.3\textwidth}
		\centering
		\includegraphics[width=\linewidth]{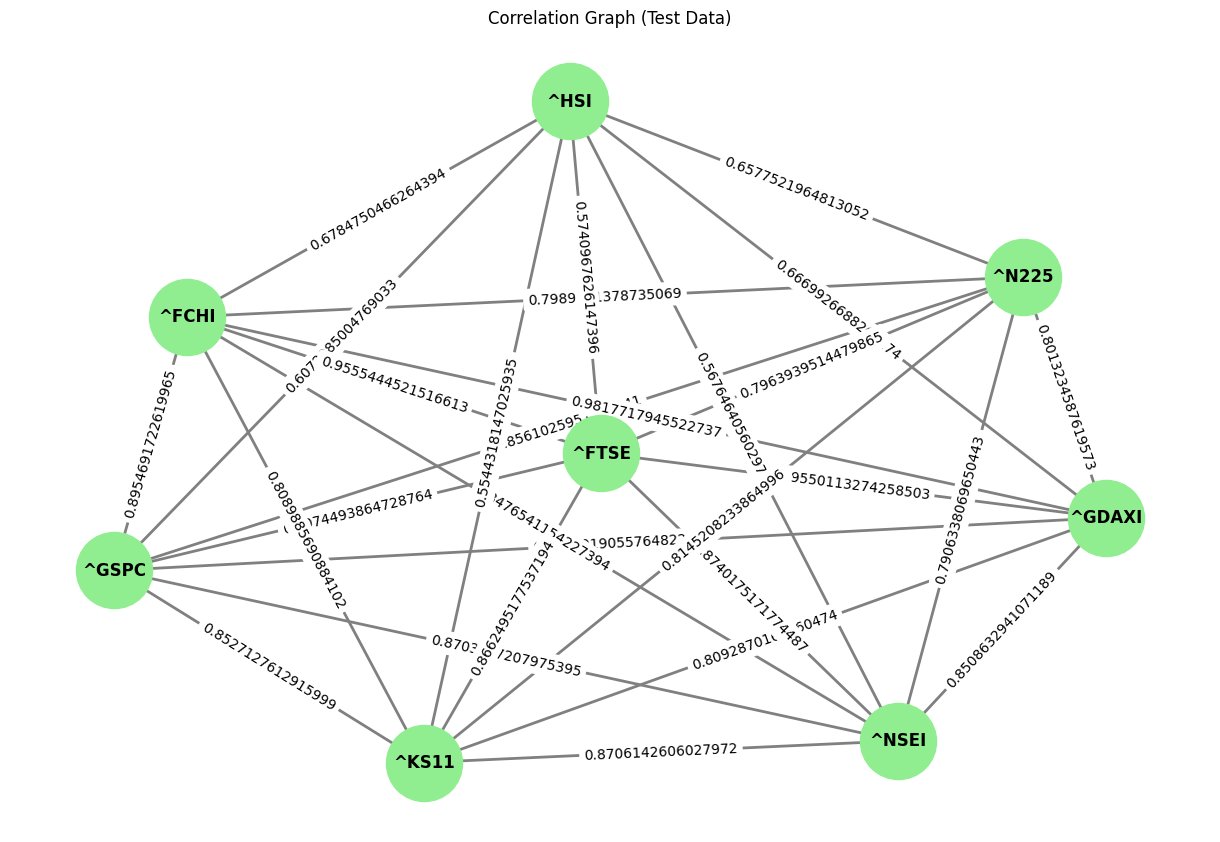}
		
	\end{subfigure}
	\caption{Visualisation of graphs by correlation method; L-R: Training, Validation, Testing}
	\label{corr_v}
\end{figure}

\noindent \textbf{Correlation Method}: In this method, Pearson correlation coefficients between the realized volatilities of the indices during the training period were calculated. These coefficients form the graph's edges, resulting in a symmetric adjacency matrix with self-loops on the diagonal (where the value is 1). The Net Correlation Index (NCI) for each market was calculated as the sum of its correlations with other markets [See Figure \ref{corr_v}]. This method helps identify the strength of the correlation between different indices, providing a clear view of the interconnectedness of these markets \citep{chordia2000commonality},\citep{karolyi2012understanding}.

\begin{figure}[H]
	\centering
	\begin{subfigure}{0.3\textwidth}
		\centering
		\includegraphics[width=\linewidth]{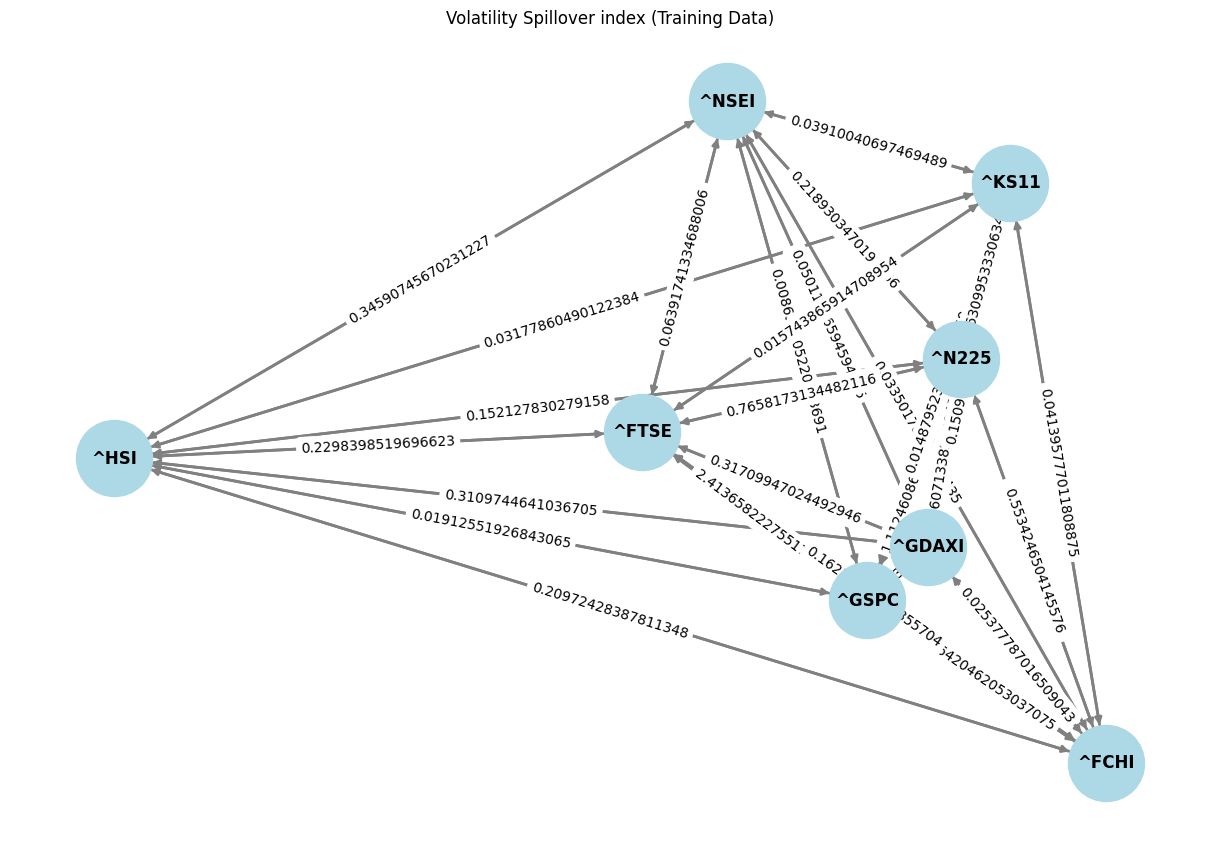}
	\end{subfigure}
	\hfill
	\begin{subfigure}{0.3\textwidth}
		\centering
		\includegraphics[width=\linewidth]{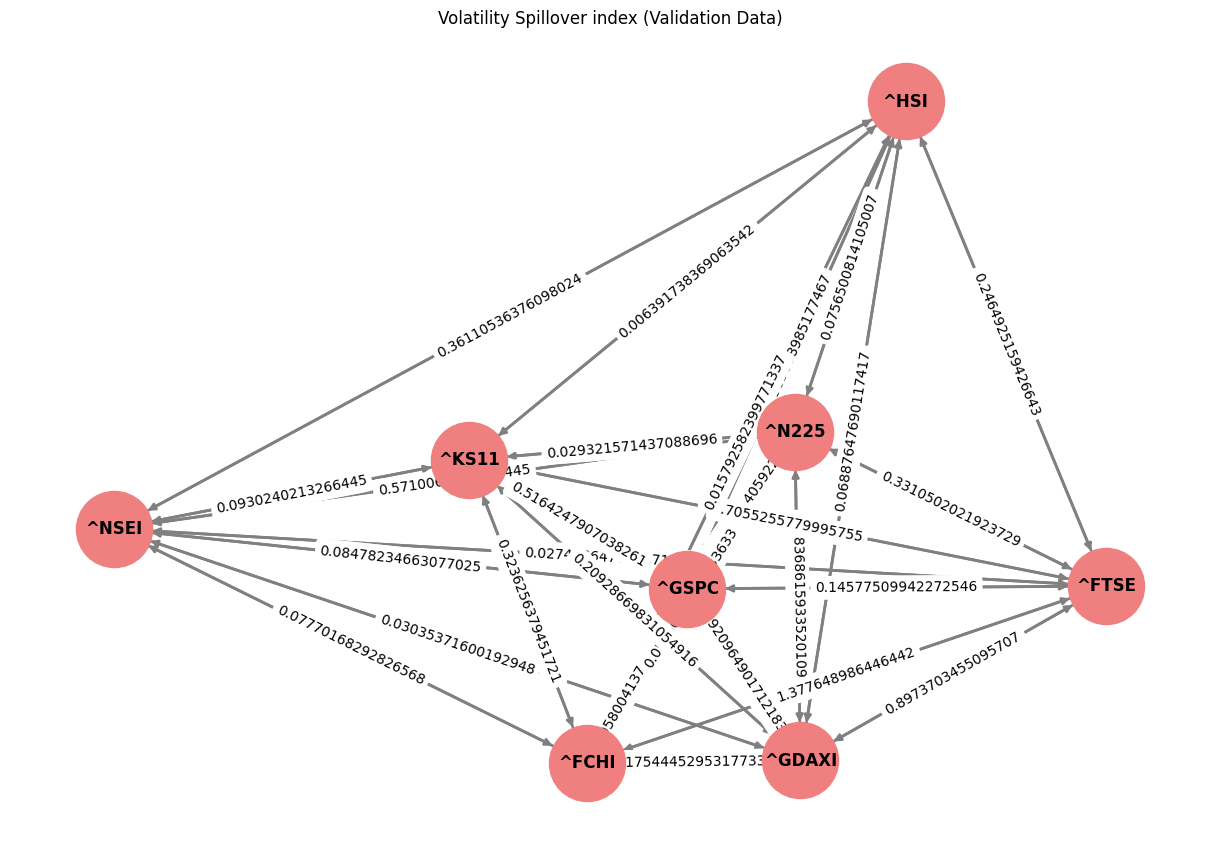}
		
	\end{subfigure}
	\hfill
	\begin{subfigure}{0.3\textwidth}
		\centering
		\includegraphics[width=\linewidth]{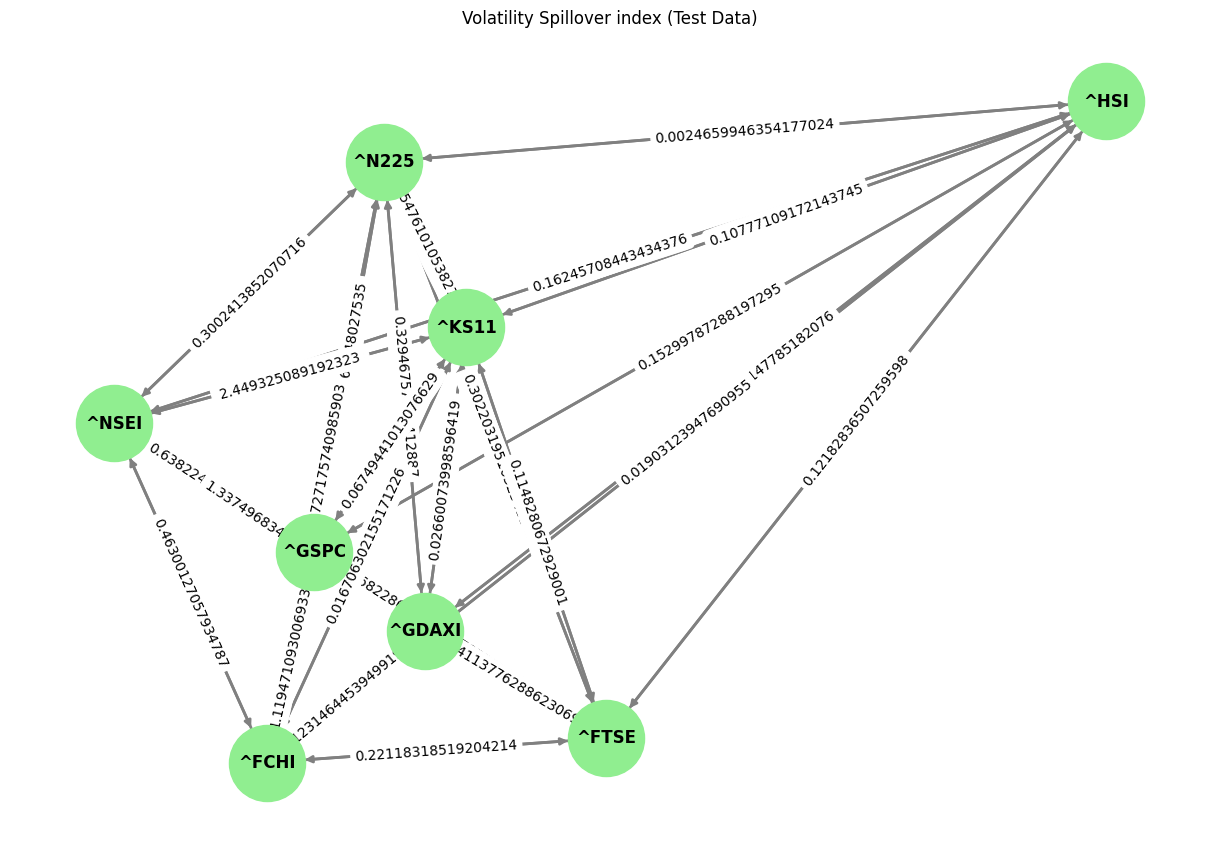}
		
	\end{subfigure}
	\caption{Visualisation of graphs by volatility spillover method; L-R: Training, Validation, Testing}
	\label{vspil_v}
\end{figure}

\noindent \textbf{Volatility Spillover Index Method}: Based on the framework by Diebold and Yilmaz (\citep{diebold2009measuring} and \citep{diebold2012better}), this method measures the degree of volatility transmission between indices using variance decomposition from a Vector Autoregressive (VAR) model. We employed a lag order of $4$ ($p = 4$) and a 5-step ahead forecast horizon ($H = 5$) to derive these indices [See Figure \ref{vspil_v}]. The resulting spillover index matrices illustrate the directional influence one market exerts over another, capturing the dynamic nature of volatility transmission in the financial markets\citep{son2023forecasting}.

\noindent Figures \ref{corr_v} and \ref{vspil_v} represent a graph where each node is a global stock market index (e.g., HSI, FTSE), and the directed edges show relationships between them. The numbers on the edges indicate the strength of the connection between two indices, which could be measured through the correlation or the volatility spillover index method. For instance, HSI has a connection strength of 0.229 (spillover method) to FTSE in the training data. The comparison between training, testing and the validation data reflects how these relationships change over time, with the numbers showing varying connection strengths.\\

\noindent Consider the training dataset in Figure \ref{corr_v}, when stock indices are highly correlated, as seen between indices like GSPC vs FCHI, HSI, KS11 (correlations of 0.91, 0.87, 0.90 respectively); KS11 vs FTSE, HSI, GSPC, GDAXI (correlations of 0.91, 0.92, 0.90, 0.90 respectively); FTSE vs GDAXI, HSI (correlations of 0.92, 0.90 respectively) and GDAXI vs FCHI (correlation of 0.96), it suggests that these markets are likely to experience volatility clustering together. If one market (e.g., GSPC) enters a period of high volatility due to some market turbulence or tensions in the USA, this volatility can impact the French market (CAC 40) because they are highly correlated. In addition, moderate to low correlations (e.g., between NSEI vs FCHI, N225 and GDAXI with correlations of 0.66, 0.67 and 0.69, respectively) indicate that while there is some level of shared volatility, these indices do not always cluster together. Thus, these cluster effects reflect interconnectedness because the regions with strong trade links, similar industry exposures, or shared investor bases tend to experience co-movements in volatility. This is especially true for global markets like the US and Europe or regional clusters like Europe or Asia.

\subsection{Model Architecture}
\begin{figure}[H]
	\centering
	\includegraphics[scale=0.55]{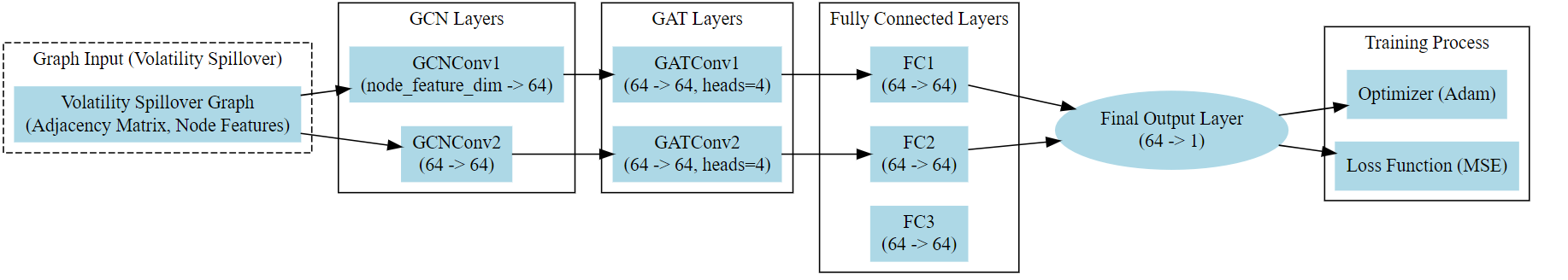}
	\caption{General design for the Temporal GAT model}
	\label{tgat}
\end{figure}

The architecture of the Temporal GAT [See Figure \ref{tgat}] is carefully designed to capture the complex dynamics inherent in financial market data. The model is structured using a combination of the GCN and GAT, which is followed by multiple fully connected layers, culminating in a predictive output layer. Each component is fine-tuned to refine the feature representation progressively, enhancing the model's predictive accuracy. The model starts with two GCN layers. The first GCN layer transforms node features from the initial dimension of the input features to a hidden dimension. This transformation is tested with hidden dimensions of 32, 64, and 128 in our implementation to determine the optimal size. The second GCN layer processes these transformed features further, applying a ReLU activation function to introduce non-linearity, which is crucial for capturing complex patterns.\\

Subsequent to the GCN layers, the architecture integrates two GAT layers that utilize attention mechanisms. These layers focus on significant nodes, prioritizing crucial information within the graph. The attention mechanism in each GAT layer is configured with multiple heads (specifically 4 or 8 heads in our tests), allowing the model to learn different aspects of the data from multiple representation subspaces simultaneously. This setup enhances the model's capacity to capture diverse relational patterns among the data points\citep{gori2005new}. The model includes three fully connected (dense) layers following the graph-based layers. Each layer maintains the hidden dimension established in prior layers and incorporates a ReLU activation function to preserve non-linear learning capabilities. These layers are crucial in synthesizing the learned graph-based features into a comprehensive form suitable for the final prediction task\citep{scarselli2008graph}. The final component of the model is a linear layer that reduces the feature dimensionality to one, matching the dimension of the target variable, such as a predicted volatility value. This layer is essential for translating the complex features learned by the model into a precise output. \\

A grid search strategy is utilized to fine-tune the model's hyperparameters, including the number of hidden dimensions, attention heads, and the learning rate. The learning rate values tested are 0.0001, 0.001, and 0.01. This optimization involves training the model across a predefined grid of parameter combinations and monitoring performance through the Mean Squared Error (MSE) on a validation set. The goal is to minimize MSE across 70 training epochs, refining the model's ability to forecast market volatility accurately.

\subsection{Other Models for Comparison}
To evaluate the performance of the Temporal GAT model (TGATM), we compared it against four alternative models, each utilizing different methodologies:
\begin{itemize}
	\item Baseline Model(BM): The BM is crafted to process the realized volatility for transforming and capturing the volatility spillover index of financial markets without integrating graph-based complexities. This model is structured as a simple multilayer perceptron (MLP), beginning with a series of fully connected layers that incrementally process features from an input dimension equivalent to the number of features per node to a specified hidden dimension. The architecture involves three hidden layers, each with dimensions configurable to be either 32, 64, or 128, allowing the model to adapt its complexity based on the richness of the input data.	
	
	\item GNN-GAT Model(GNN-GATM): The GNN-GATM leverages GNN and GAT to predict market volatility by utilizing realized volatility for data transformation. This model constructs a graph where nodes represent individual market indices and edges denote the influence exerted by one index on another, based on the Volatility Spillover Index. It starts with a GCN layer that transforms node features from an initial dimension matching the input features to a hidden dimension of 64 units. This is followed by additional GCN layers designed to enhance feature extraction, totalling 2 layers for comprehensive spatial feature learning.
	
	\item GARCH Temporal GAT Model (GARCH-TGATM): The GARCH-TGATM incorporates the GARCH methodology for transforming raw data into a format suitable for graph-based analysis, enhancing the traditional volatility modeling approach. This model leverages the strengths of both GCNs and GATs to effectively model the dynamic relationships within financial markets. The model processes node features to extract spatial features within the graph structure, starting with two GCN layers. These are then enhanced by two subsequent GAT layers that apply an attention mechanism to focus on the most influential nodes in predicting market volatility.
	
	\item Correlation Temporal GAT Model (C-TGATM): The C-TGATM leverages the structure of financial market indices, represented as nodes in a graph, with edges defined by the correlation coefficients between these indices. This model architecture uses a combination of GCN and GAT layers to capture both the local and global relational dynamics within the data.
\end{itemize}

\subsection{Model Training and Optimization}
\noindent The model training process involved early stopping to prevent overfitting and hyperparameter optimization using grid search. Key hyperparameters included:
\begin{itemize}
	\item Learning Rate: Tested at 0.0001, 0.001, and 0.01.
	
	\item Hidden Dimensions: Configurations of 32, 64, and 128 hidden units were tested.
	
	\item Attention Heads: GAT layers with 4 and 8 attention heads were evaluated.
\end{itemize}

\noindent The training was performed on Google Colab using GPUs to accelerate computations. The model's performance was evaluated using the Mean Absolute Forecast Error (MAFE) across different forecasting horizons (h = 1, 5, 10, 22) corresponding to short-term (1 day), mid-term (1 week and 2 weeks), and long-term (1 month) forecasts. The model's robustness was tested under different market conditions, including the COVID-19 crisis, which introduced significant market stress.

\section{Empirical results and discussion}\label{sec5}
\noindent This section focuses on data visualization, model analysis, sensitivity studies and robustness tests.

\subsection{Data Visualization and Analysis}
The data used in this study focuses on eight major global market indices: GSPC - S\&P 500 (USA), GDAXI - DAX (Germany), FCHI - CAC 40 (France), FTSE - FTSE 100 (UK), NSEI - Nifty 50 (India), N225 -  Nikkei 225 (Japan), KS11 - KOSPI (South Korea) and HSI - Hang Seng Index (Hong Kong). These indices were selected due to their significant influence on global financial markets \citep{son2023forecasting}. The dataset was sourced from Yahoo Finance and spans November 2007 to June 2022. The realized volatility (RV) data was computed using daily adjusted closing prices, following the approach suggested by Andersen et al.\citep{andersen2001distribution}. The dataset was divided into three subsets: a training set (1891 datasets spanning from November 2007 to August 2014), a validation set (756 datasets from September 2014 to December 2017), and a test set (1136 datasets from January 2018 to June 2022). These subsets respectively cover 50$\%,$ 20$\%,$ and 30$\%$ of the total data, ensuring a robust evaluation of the model's performance over different time periods\citep{son2023forecasting}.\\

Descriptive statistics of the realized volatility data were computed to gain insights into the underlying distribution and characteristics of the market indices. The mean values of the volatility ranged from 0.046 to 0.059, reflecting the average level of volatility across the different indices. The standard deviation, varying between 0.028 and 0.034, indicated the extent of dispersion in the volatility data. Skewness and kurtosis metrics highlighted the non-normality of the volatility distributions, with positive skewness values ranging from 2.28 to 3.23 and kurtosis values ranging from 10.47 to 20.04, suggesting the presence of heavy tails. The Augmented Dickey-Fuller (ADF) test results confirmed the stationarity of the data across all indices, with p-values significantly below 0.05. These statistics ensure that the time series data is stable and suitable for GNN modelling without further transformations\citep{barndorff2002econometric}.

\begin{table}[H]
\caption{Statistical properties of selected indices}
	\centering
	\small
	\begin{tabular}{|p{1.5cm}|p{1.2cm}|p{2cm}|p{2cm}|p{2cm}|p{1.8cm}|p{2cm}|}
		\hline
		\textbf{Ticker} & \textbf{Mean} & \textbf{Std. Deviation} & \textbf{Skewness} & \textbf{Kurtosis} & \textbf{ADF Statistic} & \textbf{ADF p-value} \\ 
		\hline
		GSPC & 0.047322 & 0.034962 & 3.144922 & 16.101063 & -5.540181 & 1.70867E-06 \\ 
		\hline
		GDAXI & 0.054093 & 0.038324 & 3.144921 & 15.566453 & -5.604956 & 1.06869E-06 \\ 
		\hline
		FCHI & 0.057695 & 0.02082 & 2.288022 & 14.471417 & -5.423957 & 3.22759E-06 \\ 
		\hline
		FTSE & 0.040868 & 0.023625 & 2.838757 & 13.337729 & -5.237459 & 7.06389E-06 \\ 
		\hline
		NSEI & 0.052276 & 0.03306 & 2.763412 & 12.677778 & -4.887057 & 3.69459E-05 \\ 
		\hline
		N225 & 0.05592 & 0.031903 & 3.234541 & 20.204222 & -5.484947 & 7.20239E-07 \\ 
		\hline
		KS11 & 0.048637 & 0.028237 & 3.098161 & 13.453776 & -4.902201 & 3.45269E-05 \\ 
		\hline
		HSI & 0.085911 & 0.033861 & 3.188015 & 18.540499 & -4.435595 & 0.0002566142 \\ 
		\hline
	\end{tabular}
	
\end{table}

\noindent To provide a visual representation of the realized volatility data across the eight global market indices, a time series plot was generated, capturing the daily adjusted closing prices. The graph below showcases the trends and fluctuations in realized volatility, highlighting high and low volatility periods. The graph illustrates the realized volatility data over the training, validation, and test periods, enabling a better understanding of market behaviours during major global events, such as the 2008 financial crisis, Brexit, and the COVID-19 pandemic, all contributing to significant market volatility.

\begin{figure}[H]
    \centering
    \begin{subfigure}{0.45\textwidth}
    	\centering
    	\includegraphics[width=\linewidth]{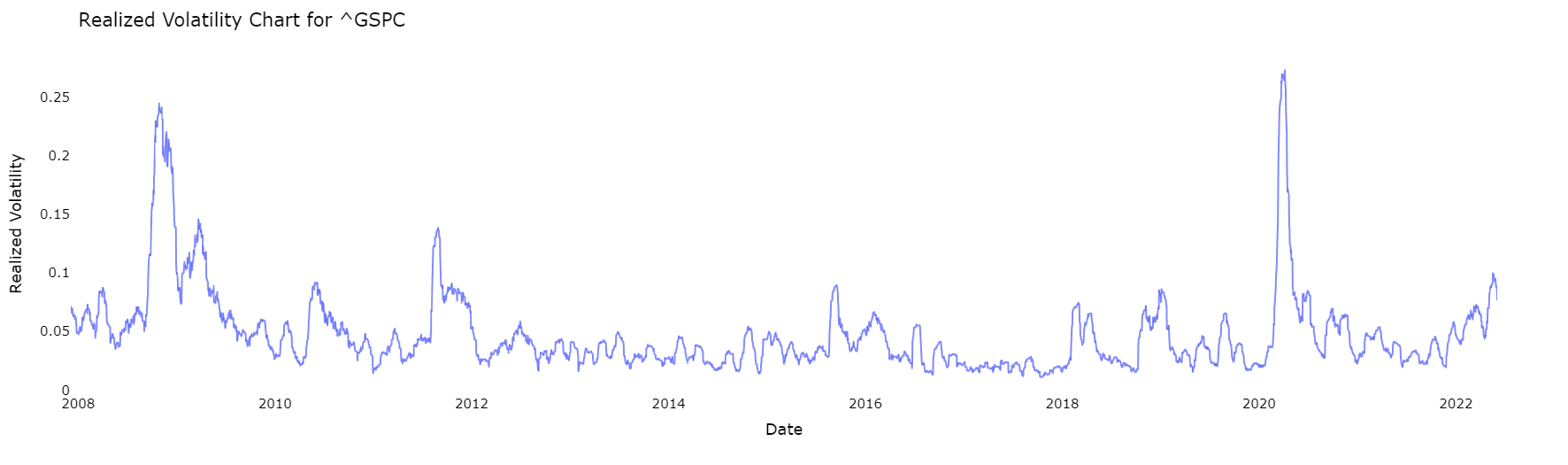}
    \end{subfigure}
	\hfill
	\begin{subfigure}{0.5\textwidth}
	\centering
	\includegraphics[width=\linewidth]{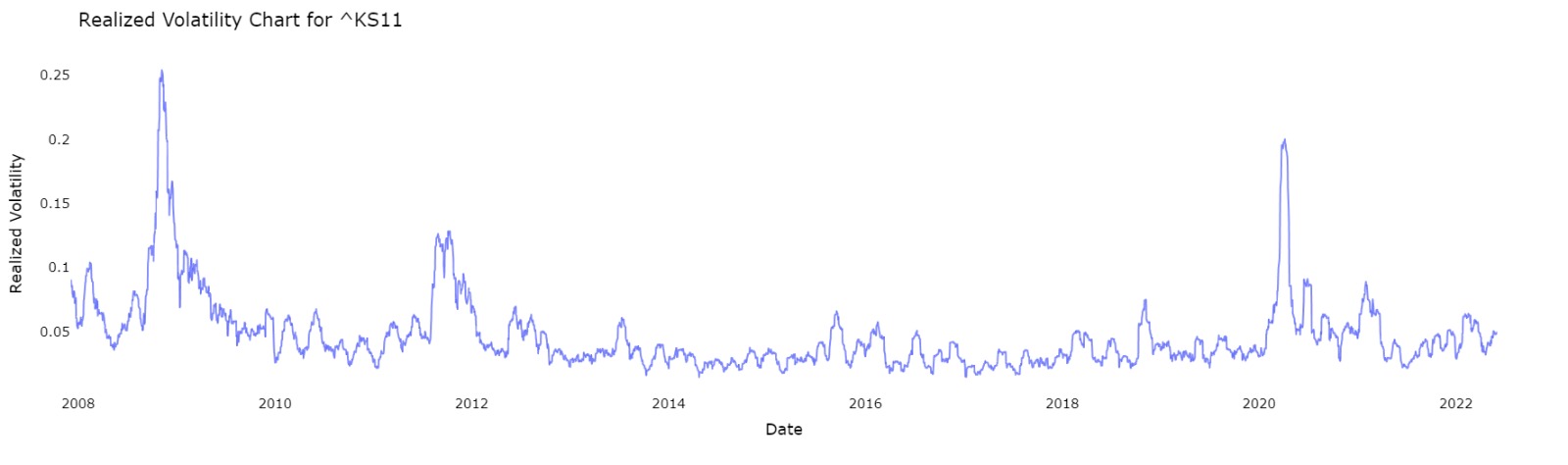}
	
	\end{subfigure}
		\begin{subfigure}{0.45\textwidth}
		\centering
		\includegraphics[width=\linewidth]{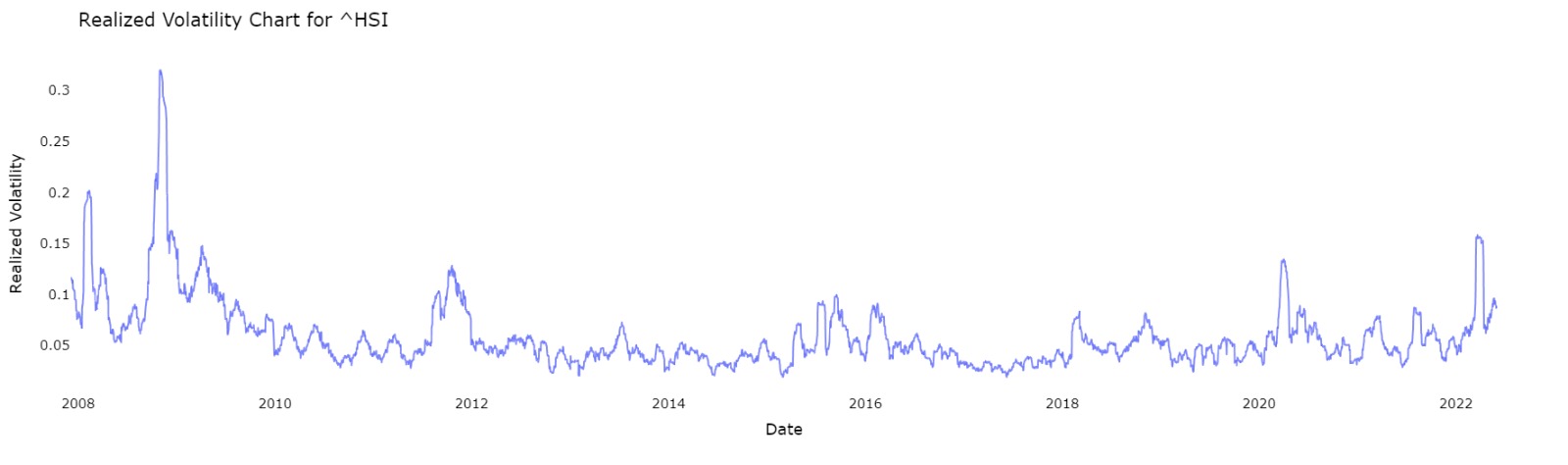}
	\end{subfigure}
	\hfill
	\begin{subfigure}{0.5\textwidth}
		\centering
		\includegraphics[width=\linewidth]{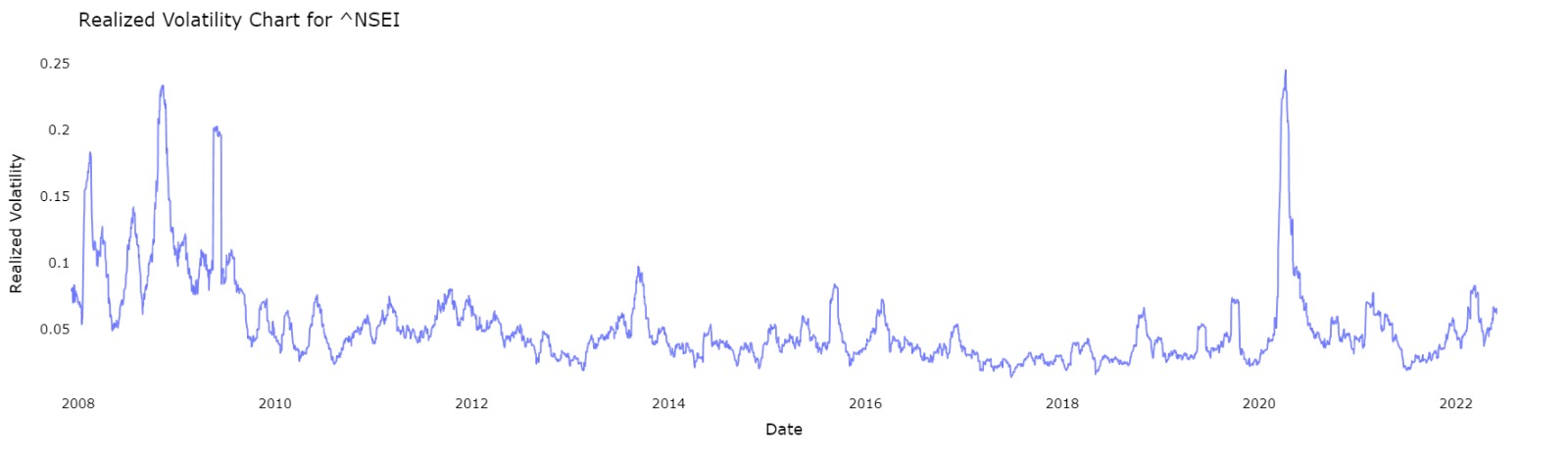}
		
	\end{subfigure}
	\begin{subfigure}{0.45\textwidth}
		\centering
		\includegraphics[width=\linewidth]{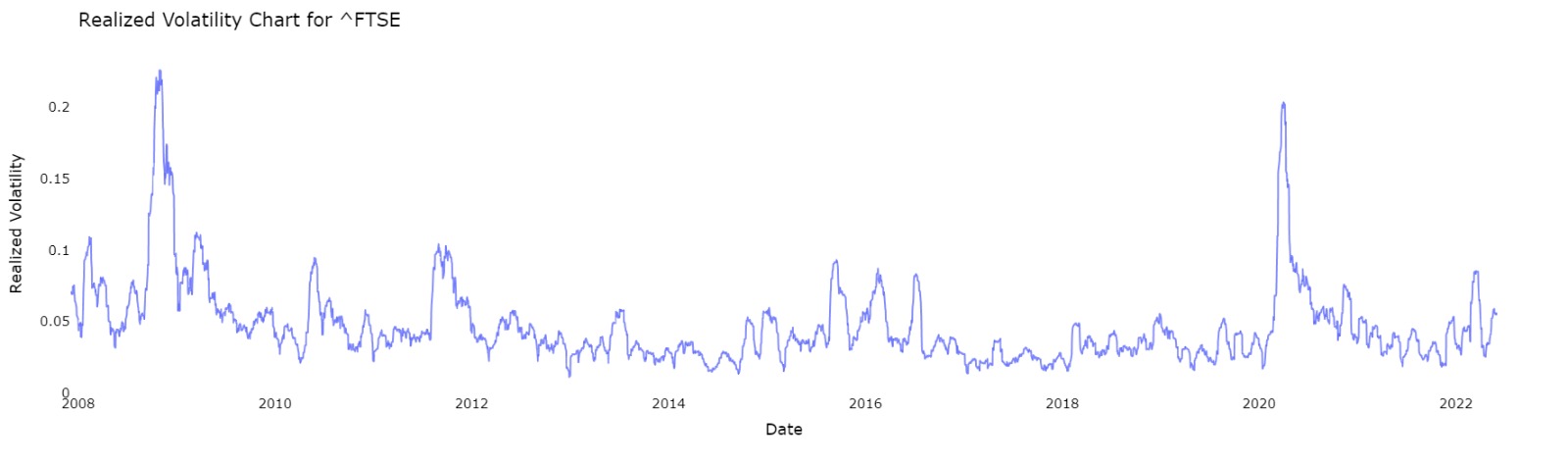}
	\end{subfigure}
	\hfill
	\begin{subfigure}{0.5\textwidth}
		\centering
		\includegraphics[width=\linewidth]{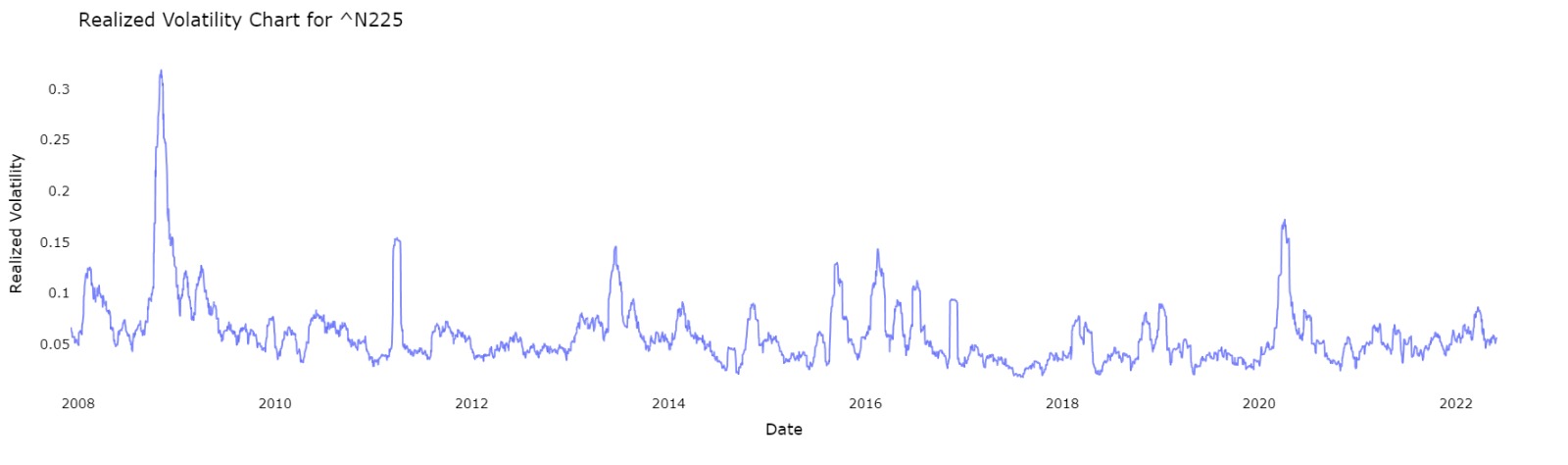}
		
	\end{subfigure}
	\begin{subfigure}{0.45\textwidth}
		\centering
		\includegraphics[width=\linewidth]{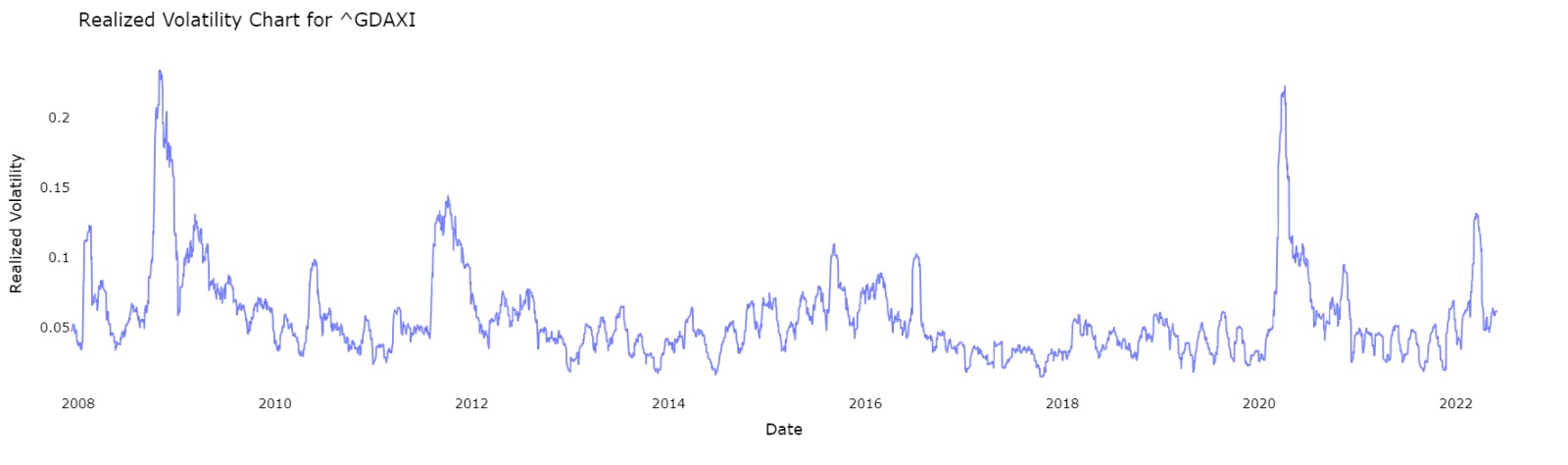}
	\end{subfigure}
	\hfill
	\begin{subfigure}{0.5\textwidth}
		\centering
		\includegraphics[width=\linewidth]{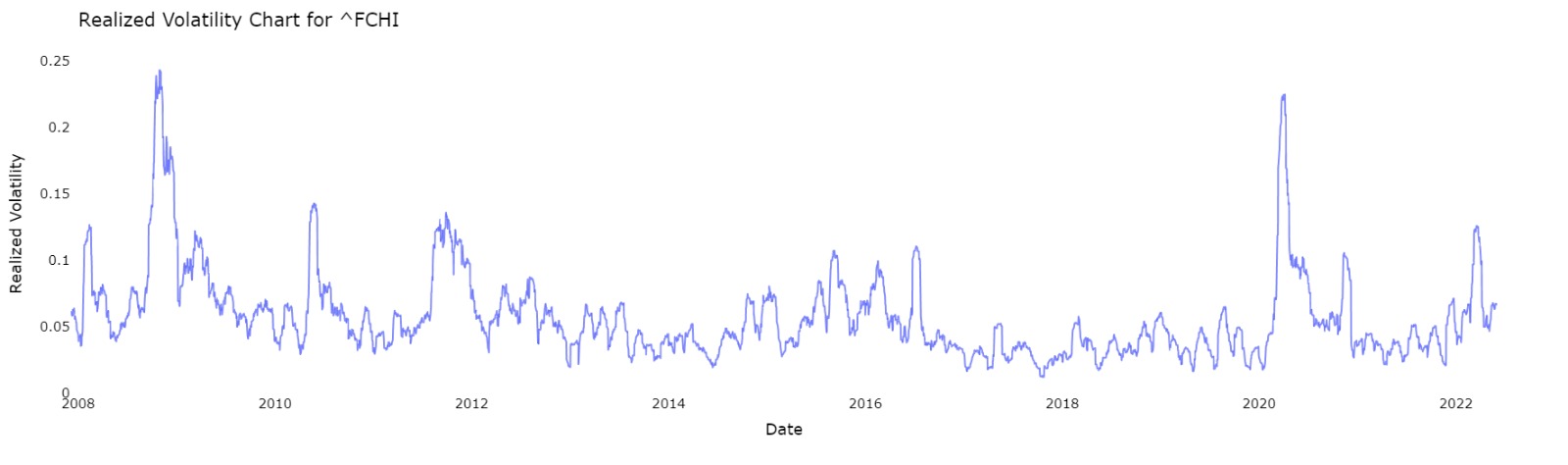}
		
	\end{subfigure}
\caption{Realized volatility for 8 selected indices [Left: GSPC, HSI, FTSE, GDAXI; Right: KS11, NSEI, N225, FCHI]}	
\end{figure}

\noindent The analysis of realized volatility data reveals significant clustering behaviour across global market indices, particularly during major financial events. 

\subsection{Comparative analysis for models}
\noindent This section compares the out-of-sample error values for the following models: BM, GNN-GATM, TGATM, GARCH-TGATM, C-TGATM and the result is presented in Table \ref{o_error}. We observe that TGATM tends to have lower MAFE values across many indices than other models, and this is especially true for GSPC (0.013323), GDAXI (0.008714), FCHI (0.008255), FTSE (0.007062), HSI (0.006439). For the MSE, TGATM has the lowest for several indices, such as FTSE (0.000067), HSI (0.000086), and GSPC (0.000215), which demonstrates good predictive accuracy. On the other hand, the GARCH-TGATM consistently has low MAFE scores for most indices, indicating its effectiveness. The model also performs well, with particularly low MSE values across indices like GSPC, GDAXI, FCHI, and NSEI. The BM and GNN-GATM consistently showed higher errors, indicating that they may not be as effective for volatility forecasting as TGATM and GARCH-TGATM. The C-TGATM has generally low MSE scores as well. For instance, GSPC has a low score of 0.000384, and HSI is 0.000284. However, indices like N225 (0.000682) show relatively higher MSE values than TGATM and GARCH-TGATM, suggesting it may lack robustness or broad applicability across all markets. However, its competitive MSE values for other indices indicate that it could still be effective in certain conditions. Thus, incorporating both temporal aspects (e.g. the evolution of stock prices over time) and structural aspects (e.g. such as how indices are related or correlated) is crucial for capturing complex patterns and improving predictive accuracy.

\begin{table}[H]
\caption{Out-of-sample error values for forecast window value of 21}
\label{o_error}
	\centering
	\begin{tabular}{|cccccc|}
		\hline
		\textbf{Indices} & \textbf{TGATM} & \textbf{BM} & \textbf{GNN-GATM} & \textbf{GARCH-TGATM} & \textbf{C-TGATM} \\\hline 
		&&&&&\\
					\multicolumn{6}{|c|}{{\bf MAFE}}\\
		\hline
		GSPC   & 0.013323     & 0.028    & 0.028  & 0.01590       & 0.01600       \\ \hline

GDAXI  & 0.008714     & 0.022       & 0.030      & 0.01363         & 0.01359       \\ \hline

FCHI   & 0.008255     & 0.023       & 0.031     & 0.01429        & 0.01426        \\ \hline

FTSE   & 0.007062     & 0.024       & 0.032     & 0.01282       & 0.01289       \\ \hline

NSEI   & 0.009622     & 0.025       & 0.033     & 0.01071        & 0.10801        \\ \hline

N225   & 0.011852     & 0.026       & 0.034     & 0.01721         & 0.01726       \\ \hline

KS11   & 0.013362     & 0.027       & 0.035     & 0.01343        & 0.01355         \\ \hline

HSI    & 0.006439     & 0.028       & 0.036     & 0.01149          & 0.01147       \\ \hline
				&&&&&\\
					\multicolumn{6}{|c|}{{\bf MSE}}\\
		\hline
		GSPC   & 0.000215     & 0.00089    & 0.0010 & 0.000885      & 0.000384     \\ \hline

GDAXI  & 0.000126     & 0.00088     & 0.0010     & 0.000482      & 0.000384     \\ \hline

FCHI   & 0.000116     & 0.00092     & 0.0011   & 0.000567       & 0.000417     \\ \hline

FTSE   & 0.000067     & 0.00096     & 0.0011    & 0.000650      & 0.000292     \\ \hline

NSEI   & 0.000169     & 0.00100       & 0.0012   & 0.000565       & 0.000257     \\ \hline

N225   & 0.000300     & 0.00104     & 0.0012    & 0.000733      & 0.000682      \\ \hline

KS11   & 0.000263     & 0.00108     & 0.0013   & 0.000736       & 0.000366     \\ \hline

HSI    & 0.000086     & 0.00112     & 0.0013    & 0.000389      & 0.000284     \\ \hline
	\end{tabular}
	\end{table}

\noindent Table \ref{o_error} considered the window size of 21 trading days. From experimental analysis, we observed that reducing the forecast window size for the TGAT will improve the short-term predictive forecasting capacity since the model will focus on short-term patterns which can be captured easily.

\begin{remark}
When constructing a financial forecast model, it is important to balance the trade-off between forecast window size and predictive performance of the model. For tasks that require short-term accuracy, like day trading, a lower  forecast window can improve performance by focusing on near-term patterns with improved reliability. On the other hand, medium- to long-term forecasting requires a larger forecast window size to capture broader trends. In the case of the TGATM, reducing the window size tends to enhance short-term forecasts, but there is a need to obtain an optimal window, since smaller window size can result to model overfitting, thereby degrading the model's performance.
\end{remark}

\subsection{Model Analysis}

\noindent In this section, we analyze the distinctions between the correlation matrix and volatility spillover index heatmaps as tools for assessing relationships among market indices. Both methodologies provide valuable insights into the interconnectedness of financial markets, but they do so through different avenues. The correlation matrix provides a snapshot of linear relationships, while the volatility spillover index heatmap offers a dynamic view of how volatility transmits between indices over time. Utilizing both tools together can provide a more comprehensive understanding of market interactions and risk dynamics. 
\begin{figure}[H]
	\centering
	\begin{subfigure}{0.45\textwidth}
		\centering
		\includegraphics[width=\linewidth]{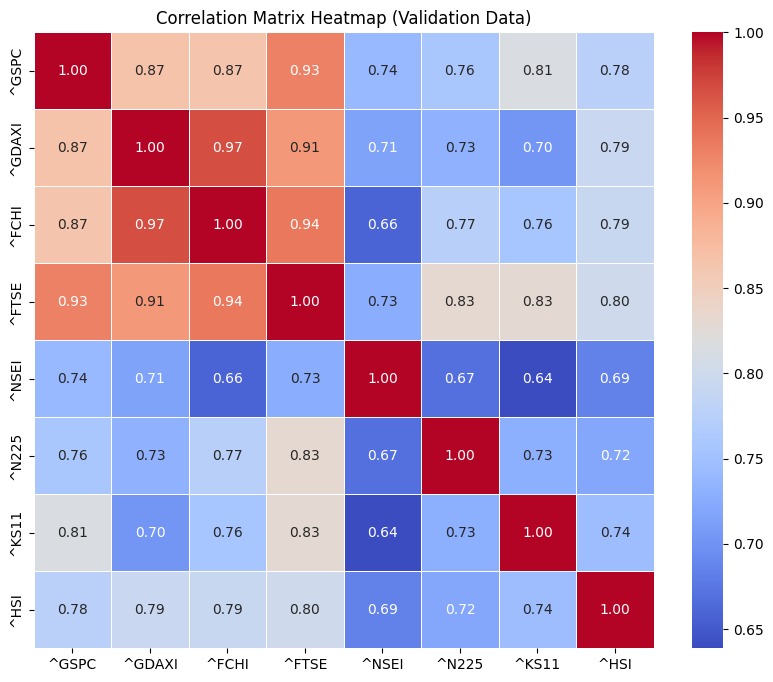}
	\end{subfigure}
	\hfill
	\begin{subfigure}{0.45\textwidth}
		\centering
		\includegraphics[width=\linewidth]{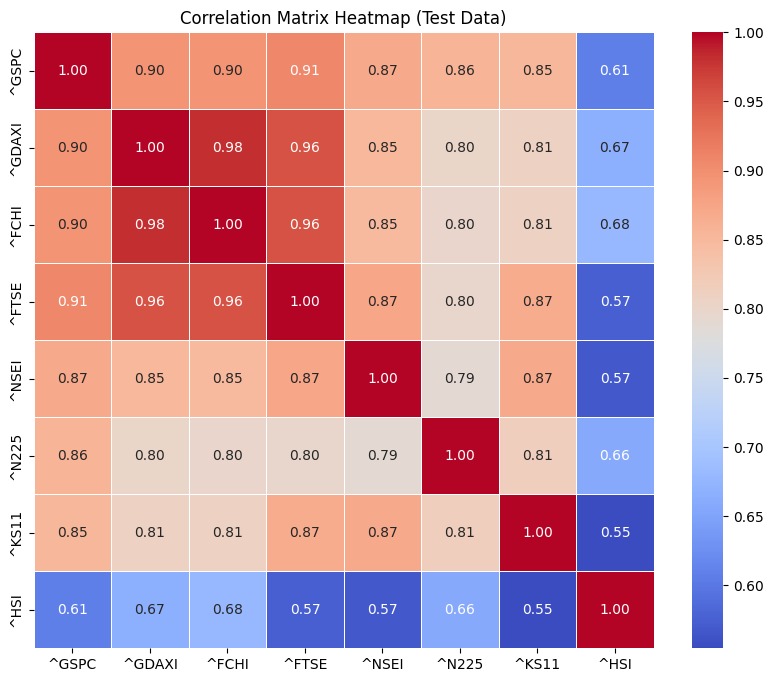}
		
	\end{subfigure}
\hfill
	\begin{subfigure}{0.45\textwidth}
		\centering
		\includegraphics[width=\linewidth]{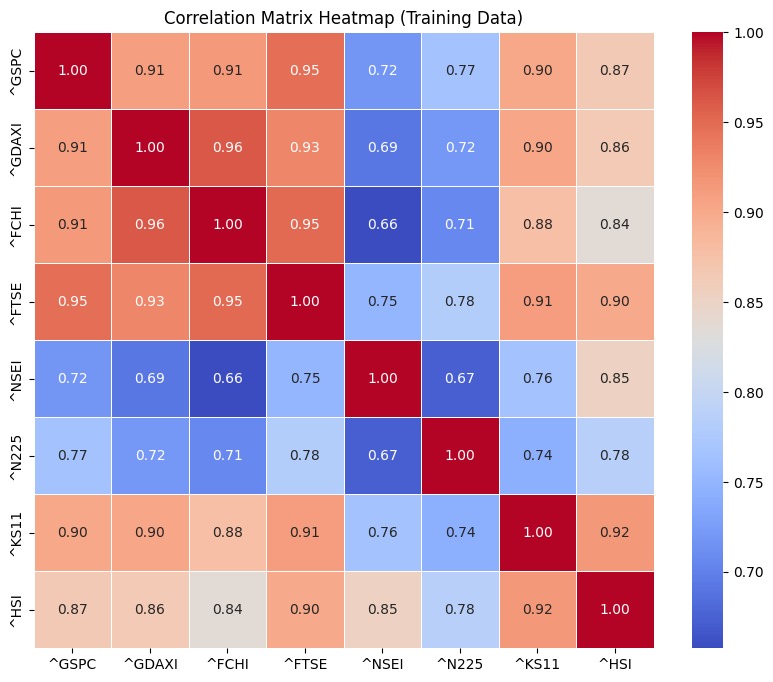}
		
	\end{subfigure}	
	\caption{Correlation index heatmaps of train, validation and test}
	\label{volsp}
\end{figure}

\noindent In Figure \ref{volsp}, we display the correlation matrix heatmaps for the training, validation and testing dataset. The figures show high interdependence across all US and European Markets datasets, with consistently strong correlations (above 0.90) between the S\&P 500 and major European indices like FTSE, GDAXI, and FCHI. These relationships indicate that Western markets move closely together, likely due to similar economic factors. For the Asian markets, the HSI shows the most independence, with lower correlations in the test data, especially with the US market. Japan (N225) shows moderate correlations, while South Korea (KS11) is more aligned with global markets. India (NSEI) shows increasing integration over time, with its correlations rising in the test data. Finally, regarding changes across datasets, we observed that training data exhibits the most robust correlations, particularly between the US, European, and South Korean markets. The validation data presents slightly lower correlations but maintains the same general trends. In contrast, the test data shows more variability, especially with HSI (Hong Kong) becoming more independent and NSEI (India) increasing its correlations with global indices.

\begin{figure}[H]
	\centering
	\begin{subfigure}{0.45\textwidth}
		\centering
		\includegraphics[width=\linewidth]{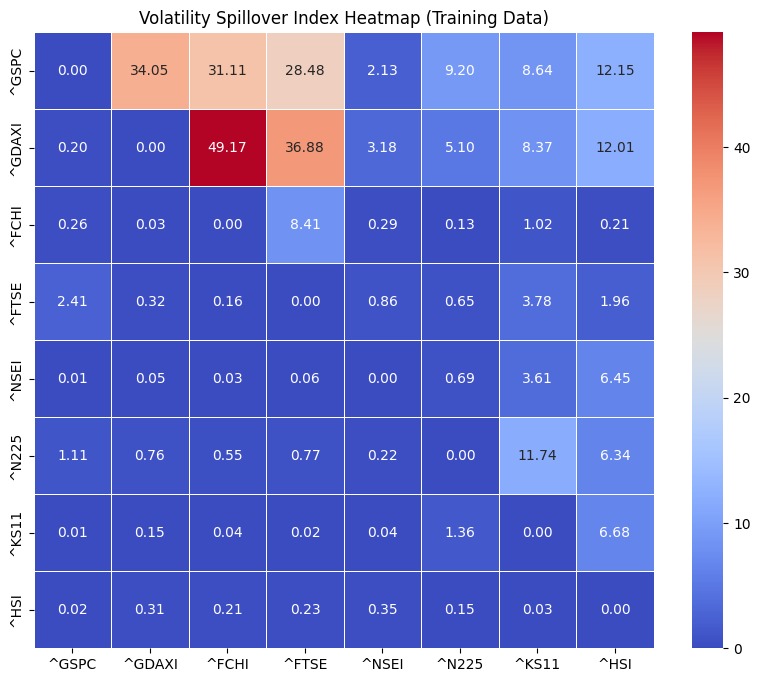}
	\end{subfigure}
	\hfill
	\begin{subfigure}{0.45\textwidth}
		\centering
		\includegraphics[width=\linewidth]{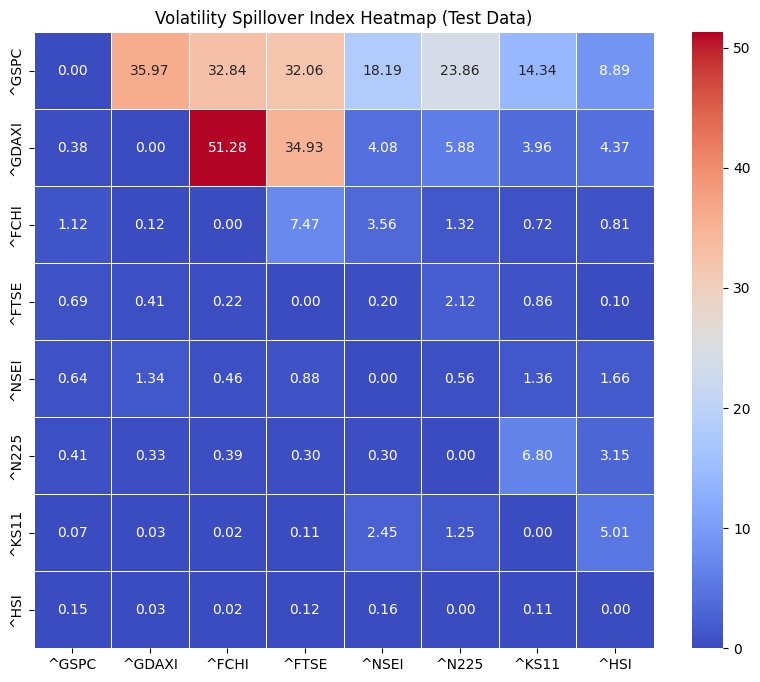}		
	\end{subfigure}
\hfill	
	\begin{subfigure}{0.45\textwidth}
		\centering
		\includegraphics[width=\linewidth]{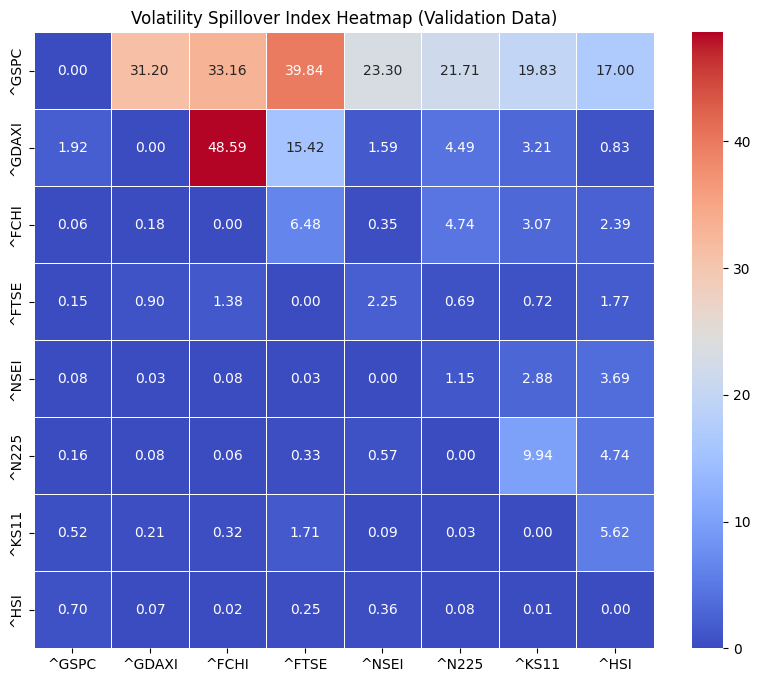}
	\end{subfigure}
	\caption{Volatility spillover index heatmaps of train, validation and test}
	\label{volspi}	
\end{figure}

\noindent In contrast, the volatility spillover index focuses specifically on the transmission of volatility between indices. Using the Diebold-Yilmaz methodology, this approach captures how shocks to one index affect the volatility of others over time. The spillover index matrix provides a directional measure of volatility transfer, illustrating which indices are net transmitters or receivers of volatility. This is particularly useful during periods of market stress, as it identifies the channels through which volatility propagates, offering a deeper understanding of market dynamics beyond mere correlation.\\

\noindent In Figure \ref{volspi}, we observe the strongest spillover (49.17) in the training dataset, which indicates that fluctuations in GDAXI significantly influence FCHI's volatility. This relationship remains robust in both the training and test datasets (51.28), suggesting a consistent dynamic between these indices. There is also a spillover value of 34.05 in the training dataset, suggesting a significant relationship, and these indicate that movements in the S\&P 500 can affect the volatility of GDAXI. Also, the spillover value remains high at 35.97 in the test dataset, further confirming the interconnectedness between these indices. The diagram also indicates some changes in the spillover dynamics, especially between FCHI and GSPC. We observed that the spillover decreased from the training data (18.19) to the test data (14.34), indicating that the influence of French market volatility on US markets may weaken. This phenomenon could result from varying market conditions or economic factors affecting each region differently over time. Finally, there are low spillover values between the HSI and the NSEI. These indices often show spillover values close to zero, indicating that they are less influenced by fluctuations in the other markets studied. For example, NSEI has several spillover values of 0.00, highlighting its independence.\\

\noindent Thus, the Temporal GAT model, when applied to volatility spillover indices, exhibits superior performance compared to models based solely on correlation analysis. The model identifies not just the relationships between indices but also how volatility evolves and impacts markets over time, leading to a more robust prediction and analysis framework. This makes the Temporal GAT model a valuable tool for understanding the complexities of financial markets, especially in periods of heightened volatility.

\subsection{Sensitivity Analysis}

The sensitivity analysis in this section focuses on evaluating the Temporal GAT model's response to varying configurations and input parameters. The aim is to assess how changes in the temporal aspects and node features impact the model's performance. We investigate three specific configurations, starting with variations in time window size to capture different levels of temporal dependencies, then evaluating the impact of different node features and concluding with analyzing the influence of hyperparameter tuning. Each of these aspects is explored in the subsections below.

\subsubsection{Temporal Aspects (Time Window Size)}
The time window size is a critical parameter determining the extent of historical data for volatility prediction. The Temporal GAT model is exposed to different temporal patterns by altering the time window size, which may capture short-term or long-term dependencies within the financial time series data. In this analysis, we analyze the trend using four time window sizes\footnote{Note: Window size of 21 represents approximately one trading month.}, 5, 15, 21, and 40 days, to investigate their effect on the prediction accuracy and model stability. Table \ref{error} below summarizes the comparative performance of different time window sizes for each metric - MAFE, MSE, RMSE, and MAPE across the eight market indices. This comparison highlights the sensitivity of the Temporal GAT model to temporal aspects and guides the selection of an optimal window size for different forecasting scenarios.

\begin{table}[H]
\caption{MAFE, MSE, RMSE and MAPE values for Different Window Sizes}
\label{error}
	\centering
	\begin{tabular}{|ccccc|}
		\hline
		\textbf{Indices} & \textbf{Window Size 5} & \textbf{Window Size 15} & \textbf{Window Size 21} & \textbf{Window Size 40} \\\hline 
		&&&&\\
					\multicolumn{5}{|c|}{{\bf MAFE}}\\
		\hline
		GSPC & 0.012568 & 0.009735 & 0.013323 & 0.014385 \\ 
		\hline
		GDAXI & 0.009990 & 0.008592 & 0.008714 & 0.012625 \\ 
		\hline
		FCHI & 0.010539 & 0.007646 & 0.008255 & 0.011707 \\ 
		\hline
		FTSE & 0.010932 & 0.005371 & 0.007062 & 0.005915 \\ 
		\hline
		NSEI & 0.009947 & 0.007371 & 0.009622 & 0.010220 \\ 
		\hline
		N225 & 0.011270 & 0.011244 & 0.011852 & 0.018505 \\ 
		\hline
		KS11 & 0.011495 & 0.009918 & 0.013362 & 0.014923 \\ 
		\hline
		HSI & 0.008951 & 0.006308 & 0.006439 & 0.008495 \\ \hline
				&&&&\\
					\multicolumn{5}{|c|}{{\bf MSE}}\\
		\hline
		GSPC & 0.000192 & 0.000124 & 0.000215 & 0.000249 \\ 
		\hline
		GDAXI & 0.000152 & 0.000123 & 0.000126 & 0.000248 \\ 
		\hline
		FCHI & 0.000168 & 0.000108 & 0.000116 & 0.000220 \\ 
		\hline
		FTSE & 0.000154 & 0.000040 & 0.000067 & 0.000052 \\ 
		\hline
		NSEI & 0.000127 & 0.000106 & 0.000169 & 0.000184 \\ 
		\hline
		N225 & 0.000244 & 0.000268 & 0.000300 & 0.000607 \\ 
		\hline
		KS11 & 0.000158 & 0.000156 & 0.000263 & 0.000349 \\ 
		\hline
		HSI & 0.000123 & 0.000081 & 0.000086 & 0.000122 \\ 		\hline
					&&&&\\
					\multicolumn{5}{|c|}{{\bf RMSE}}\\
		\hline
		GSPC & 0.013854 & 0.011133 & 0.014654 & 0.015787 \\ 
		\hline
		GDAXI & 0.012341 & 0.011079 & 0.011232 & 0.015744 \\ 
		\hline
		FCHI & 0.012975 & 0.010415 & 0.010778 & 0.014818 \\ 
		\hline
		FTSE & 0.012402 & 0.006360 & 0.008201 & 0.007222 \\ 
		\hline
		NSEI & 0.011273 & 0.010314 & 0.013000 & 0.013582 \\ 
		\hline
		N225 & 0.015627 & 0.016380 & 0.017319 & 0.024636 \\ 
		\hline
		KS11 & 0.012561 & 0.012508 & 0.016211 & 0.018693 \\ 
		\hline
		HSI & 0.011093 & 0.008983 & 0.009269 & 0.011024 \\ 		\hline
					&&&&\\
					\multicolumn{5}{|c|}{{\bf MAPE}}\\
		\hline
		GSPC & 163.77 & 50.09 & 54.71 & 39.59 \\ 
		\hline
		GDAXI & 81.60 & 21.27 & 18.58 & 16.45 \\ 
		\hline
		FCHI & 83.23 & 20.40 & 20.03 & 16.50 \\ 
		\hline
		FTSE & 104.93 & 22.98 & 25.55 & 14.62 \\ 
		\hline
		NSEI & 89.19 & 26.02 & 28.67 & 20.52 \\ 
		\hline
		N225 & 75.61 & 23.24 & 20.56 & 21.86 \\ 
		\hline
		KS11 & inf & 40.78 & 44.77 & 33.51 \\ 
		\hline
		HSI & 59.45 & 15.64 & 13.95 & 12.85 \\ 
		\hline
	\end{tabular}
	\end{table}

\noindent From Table \ref{error}, we observe that across most indices, the error metrics (MAFE, MSE, RMSE) generally increase as the window size increases. When window sizes are smaller, such as 15, FTSE performs well across a wide range of error metrics, suggesting that it is simpler to forecast with high accuracy. Regarding all metrics and window sizes, N225 typically has higher errors, especially at larger window sizes (e.g. window size 40), which increases the difficulty of accurate forecasting. In particular, \textbf{MAPE} shows a wide range of errors in GSPC, indicating that its percentage forecast error varies considerably across window sizes.\\

\noindent Furthermore, regarding the \textbf{MAFE}, we observe that the GSPC and GDAXI show increasing errors as the window size increases, and the FTSE has the lowest MAFE values, especially for smaller window sizes. For the \textbf{MSE}, the errors generally increase with larger window sizes across all indices, though some, like FTSE, maintain relatively low values throughout. For the \textbf{RMSE}, the errors grow with window size, with N225 consistently showing higher RMSE than other indices. Finally, the GSPC and KS11 demonstrate the highest variation in percentage error, while HSI and GDAXI tend to have more stable MAPE values across different window sizes when considering \textbf{MAPE}. Hence, this indicates that smaller window sizes generally lead to more accurate forecasts, whereas larger window sizes result in increased error and reduced model precision.
 
\subsubsection{Graph Properties (Node Features)}

In this section, we explore the sensitivity of the Temporal GAT model to variations in node features. Initially, the model was constructed using the 'Closing Price' as the primary node feature, representing the historical stock prices of the indices. Here, we replace the closing prices with 'Trading Volumes' to assess the impact of different node features on the model's predictive performance. The rationale behind using trading volumes is that it serves as a proxy for market liquidity and investor interest, potentially capturing different market dynamics compared to the closing prices. We evaluate the model's accuracy using the same set of performance metrics: MAFE, MSE, RMSE, and MAPE—across the same horizons, and Table \ref{node_f} below displays the result using a 1-day horizon. 

\begin{table}[H]
\centering
\caption{Comparison of MAFE, MSE, RMSE and MAPE using different node features}
\label{node_f}\scalebox{0.9}{
\begin{tabular}{|ccccccccc|}\hline
 & GSPC &  GDAXI & FCHI &  FTSE &  NSEI &  N225 &  KS11 &  HSI \\\hline
&&&&&&&&\\
\multicolumn{9}{|c|}{{\bf MAFE}}\\
Closing & 0.0133 & 0.0087 & 0.0082 & 0.0070& 0.0096& 0.0118& 0.0133& 0.0064\\
Trading Volume & 0.0281& 0.0217& 0.0210& 0.0165& 0.0661& 0.0302& 0.0219& 0.0160\\
&&&&&&&&\\
\multicolumn{9}{|c|}{{\bf MSE}}\\
Closing & 0.000215 & 0.000126 & 0.000116 & 0.000067& 0.000169& 0.000300& 0.000263& 0.000086\\
Trading Volume & 0.000941& 0.000847& 0.000856& 0.000566& 0.004514& 0.001087& 0.000559& 0.000522\\
&&&&&&&&\\
\multicolumn{9}{|c|}{{\bf RMSE}}\\
Closing & 0.0146& 0.0112 & 0.0107 & 0.0082& 0.0130& 0.0173& 0.0162& 0.0092\\
Trading Volume & 0.0306& 0.0291& 0.0292& 0.0238& 0.0671& 0.0329& 0.0236& 0.0228\\
&&&&&&&&\\
\multicolumn{9}{|c|}{{\bf MAPE}}\\
Closing & 54.70& 18.58 & 20.03 & 25.55& 28.67& 20.56& 44.77&13.95\\
Trading Volume & 129.61& 36.79& 37.18& 34.95& 192.87& 78.53& 85.52& 28.33\\\hline
\end{tabular}}

\end{table}

\noindent A comparative analysis between using the `Closing Price' and `Trading Volume' node features for a 21-day window is presented in Tables \ref{node_f}. The GSPC (S\&P 500) shows relatively higher errors for trading volume predictions across all metrics than other indices but performs moderately well regarding closing price predictions. The FTSE shows relatively lower errors for closing prices and trading volumes than other indices, as seen by the lower \textbf{MAFE} and \textbf{RMSE} values. The results indicate a noticeable change in the model's predictive accuracy, particularly in the \textbf{MAPE} values, which show significant fluctuations depending on the node features. The model using trading volumes generally produced higher error values compared to the one using closing prices, highlighting the distinct predictive value of price-based versus volume-based features. Indices such as N225 and KS11 have higher prediction errors for closing prices, indicating more volatility or complexity in forecasting these markets.\\

\noindent Regarding the \textbf{MAFE}, the values increased substantially for most indices when trading volumes were used, indicating that the model struggled to capture short-term price movements based on volume data alone. For instance, the GSPC index saw a rise in \textbf{MAFE} from 0.0133 (with closing prices) to 0.0281 (with volumes) for a 1-day horizon. For the \textbf{MSE}, similar trends were observed in \textbf{MSE}, where the trading volume-based model consistently produced higher error rates. This suggests that price data offers more granular insight into immediate price volatility. The \textbf{RMSE} values followed the same pattern, with trading volume-based predictions showing greater deviations from actual values. This deviation was particularly pronounced for longer horizons, such as the 22-day forecast. Finally, the trading volume-based \textbf{MAPE} values showed significant increases across most indices, especially for short-term horizons. For FTSE, the \textbf{MAPE} rose from 28.67 (with closing prices) to 192.87 (with volumes) for a 1-day horizon, indicating an inability to capture short-term price trends effectively. Also, the \textbf{MAPE} values, particularly for trading volume, are quite high across several indices (e.g., NSEI has a \textbf{MAPE} of 192.87), indicating significant difficulty in accurately predicting volumes in these markets. \\

\noindent Overall, the results show that the choice of node features significantly impacts model performance. While closing prices provide direct insights into market value changes, trading volumes offer a less stable foundation for short-term volatility predictions. Therefore, for the Temporal GAT model, closing prices as node features is recommended to achieve better accuracy across various horizons.

\subsubsection{Model Hyperparameters}

In this section, we explore the sensitivity of the Temporal GAT model to variations in key hyperparameters, including hidden dimensions, number of heads, and learning rates. Adjusting these parameters is crucial in determining the model's ability to generalize and capture complex relationships within the data. To identify the optimal configuration, we conducted a comprehensive grid search over a range of hyperparameter values, as described below.
\begin{enumerate}
	\item \textbf{Hidden Dimensions}: We experimented with three values for the hidden dimensions: 32, 64, and 128. The hidden dimension defines the size of the feature space in the hidden layers of the model. A larger hidden dimension allows the model to learn more complex patterns but can lead to overfitting if not appropriately regularised.
	\item \textbf{Number of Heads in GAT Layer}: We tested two values for the number of heads: 4 and 8. The number of heads controls the level of attention the model can distribute across different nodes in the graph, influencing the aggregation of information from neighbouring nodes.
	\item \textbf{Learning Rates}: We evaluated three learning rates: 0.0001, 0.001, and 0.01—to determine the optimal step size for updating the model's parameters. An appropriate learning rate ensures effective convergence during training while avoiding oscillations or premature stagnation.

\end{enumerate}

\noindent The grid search was conducted using all possible combinations of these hyperparameters, resulting in 18 configurations. Each configuration was rigorously evaluated over 70 epochs using MSE as the loss function. Each configuration's training and validation loss values were tracked to monitor convergence and identify the best-performing model. The results showed that the configuration with a hidden dimension of 64, 4 heads, and a learning rate of 0.001 achieved the lowest validation loss, indicating the best generalization capability. Table \ref{hyperp} summarizes the performance metrics for the best configuration.

\begin{table}[H]
\caption{Performance of the best model configuration}
\label{hyperp}
	\centering
	\begin{tabular}{|l|c|}
		\hline
		\textbf{Hyperparameter} & \textbf{Value} \\ 
		\hline
		Hidden Dimensions & 64 \\ 
		\hline
		Number of Heads & 4 \\ 
		\hline
		Learning Rate & 0.001 \\ 
		\hline
		Final Validation MSE & 0.000215 \\ 
		\hline
		Final Validation RMSE & 0.014654 \\ 
		\hline
		Final Validation MAPE & 54.71\% \\ 
		\hline
	\end{tabular}
\end{table}

\begin{figure}[H]
\centering
\includegraphics[scale=0.5]{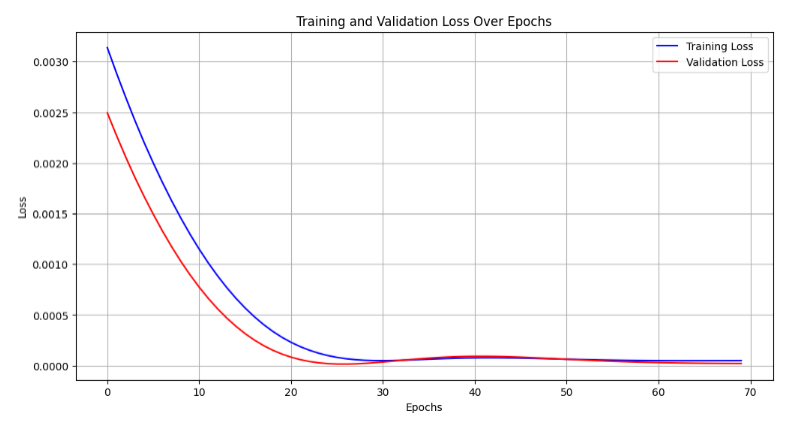}
\caption{Loss function for optimal hyperparameters}
\label{loss_f}
\end{figure}

\noindent The training and validation loss trends for the best configuration are shown in Figure \ref{loss_f}. The loss values decrease steadily over the epochs, converging to a low value without significant overfitting. This indicates that the selected hyperparameters provide a good balance between model complexity and training stability.

\noindent Overall, the hyperparameter tuning process revealed that moderate hidden dimensions and a lower number of heads in the GAT layer, combined with a learning rate of 0.001, yielded the best predictive performance. These settings will be used as the default configuration for subsequent model evaluations.

\subsection{Robustness Test: Scenario Analysis using Temporal GAT}
Robustness tests were conducted to evaluate the performance of the Temporal GAT model under different market conditions. Specifically, two distinct periods were selected: a high-volatility period (01-05-2008 to 01-09-2009) and a low-volatility period (01-04-2014 to 01-03-2016). The comparison between these periods allows for a detailed examination of how the model performs when exposed to drastically different levels of market uncertainty and price fluctuations. During the high-volatility period, the model was subjected to significant market turbulence, such as the 2008 global financial crisis. This period is characterized by abrupt changes in asset prices and increased uncertainty. As a result, the realized volatilities were significantly higher, and the prediction errors tended to increase. Table \ref{highvol} shows the error metrics for a high volatility period for each stock index in different forecast horizons.
\begin{table}[H]
\caption{Error Metrics by Horizon and Index for high volatility period}
\label{highvol}
	\centering
	\begin{tabular}{|c|c|c|c|c|c|}
		\hline
		\textbf{Horizon} & \textbf{Index} & \textbf{MAFE} & \textbf{MSE} & \textbf{RMSE} & \textbf{MAPE} \\ 
		\hline
		1 & GSPC & 0.010560 & 0.000164 & 0.012824 & 8.99\% \\ 
		\hline
		1 & GDAXI & 0.009123 & 0.000117 & 0.010810 & 8.83\% \\ 
		\hline
		5 & FCHI & 0.009927 & 0.000153 & 0.012350 & 10.07\% \\ 
		\hline
		10 & FTSE & 0.016336 & 0.000389 & 0.019730 & 19.42\% \\ 
		\hline
		22 & NSEI & 0.012764 & 0.000255 & 0.015970 & 15.27\% \\ 
		\hline
	\end{tabular}
	\end{table}

\noindent For the \textbf{Short-term Forecasts (Horizon 1)}, the MSE and the RMSE values for the high-volatility period were generally higher compared to the low-volatility period, indicating that the model struggled to capture the rapid price changes accurately. The MAPE for most indices ranged between 7.9\% and 18.8\%, reflecting the difficulty in predicting extreme price movements. FTSE and NSEI exhibited the highest prediction errors, likely due to their higher exposure to the financial crisis. For the \textbf{Medium-term Forecasts (Horizon 5 and 10)}, and with increased horizons, the prediction errors compounded, as shown by the increasing MSE and RMSE values. This suggests that while the Temporal GAT model could capture short-term fluctuations, its ability to predict long-term trends in highly volatile environments diminished. This effect is most noticeable in NSEI and KS11, where the MAPE reached 18\% and 17\%, respectively, for horizon 10. Finally, the errors became more pronounced for the \textbf{Long-term Forecasts (Horizon 22)}, with MAPE values exceeding 16\% for most indices. This outcome is expected, as long-term predictions under high-volatility conditions are inherently challenging due to the increased uncertainty in market movements.\\
	
\noindent Next, we consider the low-volatility period, spanning from 01-04-2014 to 01-03-2016, which represents a more stable market environment with less pronounced price movements. During this period, the model showed considerably better performance in terms of prediction accuracy. Table \ref{lowvol} shows the error metrics for the low volatility period for each stock index in different forecast horizons.

\begin{table}[H]
	\caption{Error Metrics by Horizon and Index for low volatility period}
\label{lowvol}
	\centering
	\begin{tabular}{|c|c|c|c|c|c|}
		\hline
		\textbf{Horizon} & \textbf{Index} & \textbf{MAFE} & \textbf{MSE} & \textbf{RMSE} & \textbf{MAPE} \\ 
		\hline
		1 & GSPC & 0.008006 & 0.000081 & 0.009014 & 28.78\% \\ 
		\hline
		1 & GDAXI & 0.024644 & 0.000693 & 0.026334 & 37.68\% \\ 
		\hline
		5 & FCHI & 0.019780 & 0.000554 & 0.023547 & 30.93\% \\ 
		\hline
		10 & FTSE & 0.004384 & 0.000027 & 0.005168 & 11.86\% \\ 
		\hline
		22 & HSI & 0.021983 & 0.000867 & 0.029452 & 30.22\% \\ 
		\hline
	\end{tabular}
\end{table}

\noindent For the \textbf{Short-term Forecasts (Horizon 1)}, the MSE and RMSE values were significantly lower compared to the high-volatility period, indicating that the model could better capture the more predictable price trends. For instance, the MAPE values for most indices ranged between 11.4\% and 37.6\%, demonstrating improved predictive capability relative to the high-volatility period. This outcome suggests that the Temporal GAT model can accurately track minor price variations when the market is stable. For the  \textbf{Medium-term and Long-term Forecasts (Horizon 5, 10, and 22)}, we observed that as the forecast horizon increased, the model maintained its robustness, with MAPE values stabilizing between 13\% and 30\% for most indices. However, there was a noticeable error increase for some indices, such as GDAXI and FCHI, which may be attributed to sporadic price shocks even in low-volatility environments. \\

\noindent In conclusion, the scenario analysis showed that the Temporal GAT model performs better under stable market conditions than turbulent ones. During high-volatility periods, the model experiences greater prediction errors due to sudden market shocks and increased uncertainty. Conversely, during low-volatility periods, the model benefits from more consistent patterns, resulting in lower MSE and MAPE values across all horizons. This comparison highlights the importance of market conditions in determining the model's efficacy and underscores the need for different modelling strategies when forecasting under varying volatility regimes.

\section{Conclusion}

This article has explored the GNN model, focusing on the Temporal GAT for predicting volatility clustering in global stock markets. The Temporal GAT effectively captures the temporal and structural dependencies in market volatility by modelling financial markets as dynamic graphs, with market indices as nodes and their relationships as edges. 
The results indicate that the Temporal GAT significantly outperforms traditional models, particularly in short- to mid-term forecasts. Its ability to dynamically assess the importance of market indices through attention mechanisms has proven essential for capturing complex relationships that other models struggle to represent. Notably, the MAFE for the Temporal GAT was markedly lower than that of baseline models, underscoring its effectiveness in volatile market conditions.\\

A key contribution of this research is the integration of volatility spillover indices with GNN architectures, which enhances the understanding of market dynamics. The volatility spillover index outperformed simpler correlation-based methods in constructing the graph, capturing the propagation of shocks across markets and providing valuable insights for financial analysts and risk managers. The leave-one-out sensitivity analysis revealed the critical roles of major global indices like the S\&P 500 and DAX in influencing market volatility. Their removal significantly reduced predictive accuracy, while indices with weaker spillover effects, like the HANG SENG, tended to add noise, indicating uneven contributions to the model's predictive power. The Temporal GAT also showed robustness during heightened market volatility, such as the COVID-19 pandemic, maintaining superior accuracy compared to traditional models. This adaptability to extreme conditions makes it a valuable tool for real-world applications where unpredictable fluctuations occur.\\

In conclusion, this research demonstrates that Graph Neural Networks, particularly the Temporal GAT, provide a robust framework for understanding and predicting volatility in global stock markets. The advancements achieved not only enhance traditional forecasting methods but also pave the way for innovations in financial modelling. As markets grow more complex and interconnected, the methodologies developed here hold significant promise for improving risk management, investment strategies, and overall decision-making in the financial industry. Despite these advancements, some limitations remain. The model's performance is sensitive to hyperparameter tuning, and the computational complexity of training GNNs on large graphs poses challenges. Additionally, assuming static graph structures during training may not fully capture the evolving nature of financial markets during structural shifts.\\

Future research could expand the Temporal GAT model's application to other asset classes, such as bonds, commodities, or cryptocurrencies, yielding further insights. Incorporating advanced features like macroeconomic indicators, sentiment analysis, or high-frequency trading data could enhance predictive capabilities. Exploring dynamic graph architectures, such as temporal graph networks that adapt to changing market relationships, may improve the model's effectiveness in capturing evolving conditions.

\section*{Funding} 
\noindent This research received no external funding.
\section*{Availability of data, code and materials} 
\noindent Please contact the corresponding author for request.
\section*{Contributions} 
\noindent All authors contributed equally to the paper. All authors read and approved the final manuscript.
\section*{Declarations}
\noindent\textbf{Conflict of interest}: All authors declare that they have no conflict of interest.\\
\textbf{Ethical approval}: This article does not contain any studies with human participants or animals performed by any of the authors.

\addcontentsline{toc}{section}{\bf References}
\bibliographystyle{agsm}
\bibliography{references2}

\end{document}